\documentclass[11pt]{article}
\usepackage{jheppub1}
\usepackage{amsmath,amssymb,amsfonts,graphicx,subfigure}

\def\bea{\begin{eqnarray}}
\def\eea{\end{eqnarray}}
\def\nn{\nonumber}
\def\ba{\begin{array}}
\def\ea{\end{array}}
\def\nn{\nonumber}
\def\Tr{\text{Tr}}
\def\sgn{\text{sgn}}
\def\J{\mathcal{J}}
\def\V{\mathcal{V}}
\def\S{\mathcal{S}}

\def\O{\mathcal{O}}

\def\Pf{\text{Pf}}

\title{Note on entropy dynamics in the Brownian SYK model}

\author[1]{Shao-Kai Jian,}

\author[2,1]{Brian Swingle}

\affiliation[1]{Condensed Matter Theory Center and Joint Quantum Institute, Department of Physics, University of Maryland, College Park, Maryland 20742, USA}

\affiliation[2]{Department of Physics, Brandeis University, Waltham, Massachusetts 02453, USA}

\emailAdd{skjian@umd.edu, bswingle@umd.edu}

\abstract{We study the time evolution of R\'enyi entropy in a system of two coupled Brownian SYK clusters evolving from an initial product state. The R\'enyi entropy of one cluster grows linearly and then saturates to the coarse grained entropy. This Page curve is obtained by two different methods, a path integral saddle point analysis and an operator dynamics analysis. Using the Brownian character of the dynamics, we derive a master equation which controls the operator dynamics and gives the Page curve for purity. Insight into the physics of this complicated master equation is provided by a complementary path integral method: replica diagonal and non-diagonal saddles are responsible for the linear growth and saturation of R\'enyi entropy, respectively.}

\begin{document}
\maketitle

\newpage
\parskip=10pt

\section{Introduction}

Interest in the entropy dynamics of quantum many-body systems has increased dramatically in the past few years. In the context of the black hole information problem~\cite{Hawking:1974particle}, recent works have derived the Page curve~\cite{Page:1993average} of an evaporating black hole from holographic calculations~\cite{Penington:2019entanglement,Almheiri:2019the,Penington:2019replica,Almheiri:2019replica}. These calculations have been quickly generalized to various related situations~\cite{Almheiri:2020page, Almheiri:2019islands, Almheiri:2019entanglement, Rozali:2019information, Gautason:2020page, Hashimoto:2020islands, Hartman:2020flat, Hollowood:2020islands,Krishnan:2020Page, Chen:2020evaporating, Chen:2020bra, Hartman:2020islands, Anegawa:2020notes,Akers:2020leading, Balasubramanian:2020islands, Balasubramanian:2020entanglement, Ling:2020island, Bhattacharya:2020topological, Marolf:2020observation}. In particular, using a semiclassical saddle point analysis of the gravitational path integral, it was shown that replica wormhole configurations, though exponentially suppressed, become dominant after the Page time~\cite{Penington:2019replica, Almheiri:2019replica}. Meanwhile, entropy dynamics are also much explored in various circuit models~\cite{Dahlsten:2007emergence,Znidaric:2007exact, Lashkari:2011towards,Nahum:2016quantum, vonKeyserlingk:2017operator,Nahum:2018operator,Rakovszky:2018diffusive, Vedika2018:operator,Chan:2018solution,Chan:2018spectral, Bertini:2018exact, Zhou:2019operator, Zhuang:2019scrambling}. Notably, Page-like behavior has been obtained in random circuit models~\cite{Czech:2011black, Mathur:2011correlations, Bradler:2015one,  Tokusumi:2018quantum, Piroli:2020a, Liu:2020a}, including the Brownian SYK model~\cite{Sunderhauf:2019quantum}. It is thus interesting to make a connection between these two types of methods in a single model.

Since the Brownian SYK model is amenable to both saddle point methods and circuit techniques, in this note we report a calculation of the Page curve in a model consisting of two coupled Brownian SYK clusters using both saddle point methods and operator dynamics. The quantity we are interested in is the R\'eny entropy of one cluster after tracing out the other. To formulate a path integral representation of the R\'enyi entropy, the initial state is taken to be a tensor product of thermofield double (TFD) states in each subsystem obtained by doubling the Hilbert space to left and right sides\footnote{The Brownian model does not have a fixed Hamiltonian with which we can define a finite temperature, so we consider the infinite-temperature TFD state in each subsystem, i.e., a maximally entangled state.}. This is the Brownian version of the setup in~\cite{Penington:2019replica}: the entanglement between left and right sides is maximal, whilst it is initially zero between the two subsystems. We find analytic solutions to the saddle point equations and show that replica diagonal and non-diagonal solutions are responsible for early time growth and the late time saturation of the R\'enyi entropy, respectively. 

On the other hand, the density matrix dynamics can be directly analyzed using an operator dynamics approach. For simplicity we take a tensor product of Kourkoulou-Maldacena states~\cite{Kourkoulou:2017pure} in each subsystem as initial state. Again, there is no entanglement between the two subsystems initially. Then the system is evolved under the full Hamiltonian, and the Page curve is obtained from a corresponding master equation. We compare the results from these two methods, and find excellent agreement even though the two approaches use slight different initial states. Complementing the saddle point method, the master equation knows about the microstate from the perspective of symmetry, i.e., the Fermi parity in our case. The Hilbert space factorizes into different Fermi parity sectors, leading to an order one correction to the coarse grained entropy. Note that while the operator dynamics approach gives access only to the second R\'enyi entropy, in the saddle point method, we are able to get solutions for R\'enyi entropy for arbitrary R\'enyi index $n$. 

The rest of the paper is organized as follows. In Section~\ref{sec:saddle} we discuss a saddle point analysis of the path integral representation of R\'enyi entropy. Both replica diagonal and replica non-diagonal solutions are obtained analytically and checked numerically. The Page curve is obtained using these solutions, with the replica diagonal solution being responsible for the linear growth and the replica non-diagonal solution leading to the saturation to the coarse grained entropy. In section~\ref{sec:master} we study the Page curve using the operator dynamics of the Brownian SYK model. We derive the master equation governing the operator size distribution function. The initial linear growth and late time saturation can be obtained analytically from the master equation, which shows exact agreement with the saddle point analysis.

\section{R\'enyi entropy dynamics from saddle points} \label{sec:saddle}

\subsection{Coupled Brownian SYK clusters}

The time-dependent Hamiltonian of two coupled Brownian SYK clusters labelled by $a=1,2$ is given by
\bea\label{eq:Hamiltonian}
	H(t) = \sum_{|A|=q, a=1,2} J_{A}^a(t) [\psi_a]_A + \sum_{|A|=|B|=q/2} V_{A,B}(t) [\psi_1]_A [\psi_2]_B,
\eea
where $A=j_1...j_{|A|}$ denotes an ascending list of length $|A|$, $q$ is an even integer, and
\bea
	[\psi_a]_A \equiv i^{|A|/2} \psi_{j_1,a} \psi_{j_2,a} ... \psi_{j_{|A|},a},
\eea
is a short-hand notation for an $|A|$-body interaction. The $\psi_{j,a}$, $j=1,...,N_a$, are Majorana fermions in subsystem $a$, and satisfy $\{\psi_{j,a}, \psi_{j',a'}\} = \delta_{jj'}\delta_{aa'}$. The summations in Eq.~\eqref{eq:Hamiltonian} are over all possible lists with the indicated number of fermions. $J_{A}^a(t)$ and $V_{A,B}(t)$ are Brownian random interactions within and between the two subsystems, respectively. The interaction strength is drawn from a Gaussian distribution with mean zero and variance given by
\bea
	\overline{J_{A}^a(t)J_{A'}^{a'}(t')} &=& \frac{2^{q-1} q! }{q^2 N_a^{q-1}} \J \delta(t-t') \delta_{A,A'} \delta_{a,a'}, \\
	\overline{V_{A,B}(t)V_{A',B'}(t')} &=& \frac{2^{q} (q/2)!^2}{q^2 N_1^{(q-1)/2} N_2^{(q-1)/2} } \V \delta(t-t') \delta_{A,A'} \delta_{B,B'}, \\
	\delta_{A,A'} &\equiv& \delta_{j_1, j_1'} ... \delta_{j_{|A|}, j_{|A'|}'}.
\eea
The over line denotes an average over the Gaussian distribution of couplings. The interaction strength has dimension one (the dimension of energy), so $\J$ and $\V$ also have dimension one, while the $\delta$-function makes up another dimension one and also indicates that the couplings are Brownian variables, uncorrelated in time. Regarding the prefactor, the dependence on $N_a$ is chosen to facilitate the large-$N$ limit and the dependence on $q$ is chosen to facilitate the large-$q$ expansion in Appendix~\ref{append:twist}. In general, the coupling between two subsystem does not have to be the same $q$-body interaction as the interaction within each subsystems, but we make such a choice for simplicity.

\subsection{Setup}

To investigate the entropy dynamics, we consider a similar setup to Ref.~\cite{Penington:2019replica}: starting from the tensor product of two thermofield double (TFD) states in each of the subsystems ($a=1,2$), we focus on the R\'enyi entropy of subsystem $a=1$ by tracing out subsystem $a=2$. Because Brownian random interactions do not conserve energy, we simply consider an infinite temperature TFD state, which is a maximally entangled state. To prepare such a state, we double the Hilbert space by introducting left ($L$) and right ($R$) copies of the fermions, $\psi_{j,a,L}$ and $\psi_{j,a,R}$, for both subsystems $a=1,2$. Then the maximally entangled state and the initial density matrix are given by
\bea
    (\psi_{j,a,L} + i \psi_{j,a,R} )|\infty \rangle =0, \quad \forall a=1,2, \quad \forall j=1,...,N, \quad \rho_0 = |\infty \rangle  \langle \infty |.
\eea

Consider a time evolution generated by the sum of left and right Hamiltonians. The random couplings are identical between the two sides, up to an overall coefficient, with $H_L(t) = H(t;\psi_{j,a,L})$ and $H_R(t) = (-1)^{q/2} H(t; \psi_{j,a,R})$. This choice implies that $H_R |\infty\rangle = H_L |\infty \rangle$. Hence, the reduced density matrix $\rho_1$ of the subsystem $a=1$ (including both $L$ and $R$ pieces) at time $t$ is
\bea\label{eq:rho_t}
    \rho(t) = U(t) \rho_0 U^\dag(t), \quad U(t) = \mathcal{T} e^{-i \int_0^t dt' (H_L(t')+H_R(t'))},  \quad \rho_1(t) = \Tr_2[ \rho(t) ].
\eea
where $\mathcal T$ denotes time ordering, and $\Tr_a$ denotes the trace over subsystem $a$. The $n$-th R\'enyi entropy is $e^{-(n-1) S_n} = \Tr_1[ \rho_1(t)^n]$. This joint left-right evolution is equivalent to a single sided evolution for twice the time, i.e., $U(t) = \mathcal{T} e^{-i\int_0^{2t} dt' H(t')}$. 

\begin{figure}
    \centering
    \includegraphics[width=6cm]{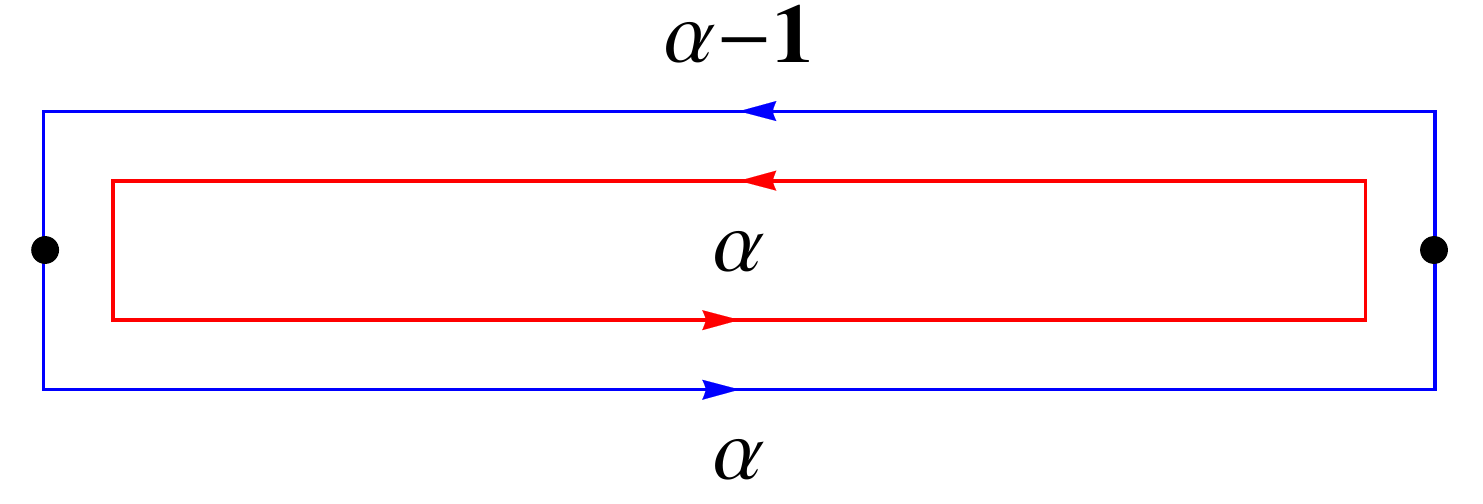}
    \caption{The schematic plot of the Keldysh contour and twist operator. The blue and red contours represent subsystem $a=1$ and $a=2$, respectively. The black dots at two ends of contour represent twist operators. The red contour is the $\alpha$ replica while in $s=-$ of the blue contour it changes to $\alpha-1$ due to the twist operator. \label{fig:contour}}
\end{figure}

A path integral representation of the trace of the $n$-th power of the reduced density matrix is obtained via a standard replica trick using $n$ copies of the system and twist fields to implement modified boundary conditions on subsystem $1$. This formulation gives $\Tr_1[\rho_1^n] = \frac{Z_{(n)}}{Z_{(1)}^n}$, where $Z_{(n)}$, $n \ge 2$ is the replicated partition function with twist operators inserted, and $Z_{(1)} \equiv Z$ is the partition function of a single replica. The replicated partition function $Z_{(n)} = \int [D\psi] e^{-I}$ can be implemented in a Keldysh contour with two twist operators at $t=0$ and $t=T$, respectively~\cite{Penington:2019replica}. The insertion of twist operators in the contour is shown in Fig.~\ref{fig:contour}. The effective action for the path integral of the replicated systems is
\bea
    I =  \sum_{s=\pm} s \int_0^T dt \Big( \frac12 \sum_{j,a,\alpha} \psi_{j,a,s}^\alpha \partial_t \psi_{j,a,s}^\alpha + i \big( \sum_{A,a,\alpha} J_A^a [\psi_{a,s}^\alpha]_A + \sum_{A,B,\alpha,\beta} g_s^{\alpha\beta} V_{A,B} [\psi_{1,s}^\alpha]_A [\psi_{2,s}^\beta]_B  \big) \Big),
\eea
where $s = \pm $ stands for the forward and backward contour, $\alpha, \beta = 1,2,...,n$ are the replica indices, and $g^{\alpha\beta}_{+} = \delta^{\alpha\beta}$, $g^{\alpha\beta}_{-} = \delta^{\alpha+1,\beta} \equiv \epsilon^{\alpha\beta}$ is due to the twist operator.

At this point, the Brownian random interactions are integrated out. Strictly speaking, what we calculate is the logarithm of the average of the replicated partition function, i.e., $\frac1{1-n} \log \overline{\Tr_1[\rho_1^n]} $, instead of the average R\'enyi entropy $\frac1{1-n} \overline{\log \Tr_1[\rho_1^n]}$ which involves an additional replica trick to obtain the averaged logarithm. Nevertheless, due to the large-$N$ structure of the Brownian models, we expect the circuit-to-circuit fluctuation are suppressed~\cite{vonKeyserlingk:2017operator, Zhou:2018emergent} so that both quantities agree with each other at large $N$. After integrating out the Brownian variables and introducing the bilocal fields $G$ and $\Sigma$,
\bea
	&&\int d\hat G_a \delta\Big( G_{a,ss'}^{\alpha\beta}(t_1,t_2) - \frac1{N_a} \sum_j \psi_{j,a,s}^\alpha(t_1) \psi_{j,a,s'}^\beta(t_2) \Big) \\
	&& =\int d\hat G_a d \hat \Sigma_a \exp \Big[ -\frac{N_a}2 \Sigma_{a,ss'}^{\alpha\beta}(t_1,t_2) \Big( G_{a,ss'}^{\alpha\beta}(t_1,t_2) - \frac1{N_a} \sum_j \psi_{j,a,s}^\alpha(t_1) \psi_{j,a,s'}^\beta(t_2) \Big) \Big],
\eea
we arrive at the following effective action,
\bea \label{eq:action}
	- I &=& \sum_a N_a \Big[ \log \Pf [ \partial_t \hat \sigma^z - \hat \Sigma_a] + \int dt_1 dt_2  \big( - \frac{1}2  \Sigma_{a,ss'}^{\alpha\beta} G_{a,ss'}^{\alpha\beta} + \frac{\J }{4 q^2} \delta(t_{12}) c_{ss'} (2G_{a,ss'}^{\alpha\beta}(t_1,t_2))^q \big) \Big] \nn\\
	&& + \sqrt{N_1 N_2} \frac{\V}{2q^2} \int dt_1 dt_2 \delta(t_{12}) c_{ss'}(2G_{1,ss'}^{\alpha\beta}(t_1,t_2))^{q/2} g^{\alpha\gamma}_s g^{\beta\delta}_{s'} (2G_{2,ss'}^{\gamma\delta}(t_1,t_2))^{q/2}.
\eea
where $t_{12} \equiv t_1 - t_2$, and $c_{++}=c_{--} = - 1$, $c_{+-} = c_{-+} =  1$ is due to the Keldysh evolution. 
The summation over the replica indices and the contour indices is implicit.
$\hat \sigma^z$ denotes the Pauli matrix acting on contour space.

From the above effective action the Schwinger-Dyson equations are
\bea
\label{eq:SD1}	\hat G^{-1}_a &=& \hat \sigma^z \partial_t - \hat \Sigma_a, \\
\label{eq:SD2}	\Sigma_{a,ss'}^{\alpha\beta} &=& c_{ss'}\delta(t_{12}) \Big[   \frac{\J}{q}   (2G_{a,ss'}^{\alpha\beta})^{q-1} +  \sum_{\gamma\delta}\sqrt{\frac{ N_{\bar a}}{N_a}} \frac{\V}{q} (2G_{a,ss'}^{\alpha\beta})^{q/2-1} [g_{(a)}]^{\alpha\gamma}_s [g_{(a)}]^{\beta\delta}_{s'} (2G_{\bar a,ss'}^{\gamma\delta})^{q/2} \Big], \nn \\
\eea
where we have defined $g_{(1)}=g$, $g_{(2)}= g^T$, and $\bar 1 =2 $, $\bar 2 = 1$.

\subsection{Saddle point solutions}

For simplicity, we assume the two clusters have equal numbers of Majorana fermions $ N_1 = N_2 = N$. To look for a replica diagonal solution, we can start by looking for a solution when the inter-cluster coupling $\V=0$. In this case, the problem reduces to $n$ independent replicas. Moreover, because of the Brownian nature of the problem, the self-energy is local in time~(\ref{eq:SD2}). Starting from the Green's functions ansatz 
\begin{equation}
G_{a}^{\alpha\beta} (t_1,t_2) = \delta^{\alpha\beta} \frac{f(t_{12}) }2 \left( \ba{cccc} \sgn(t_{12}) & -1 \\ 1 & -\sgn(t_{12})   \ea \right),
\end{equation}
one uses~(\ref{eq:SD2}) to get 
\bea
	\Sigma_{a}^{\alpha\beta}(t_1,t_2) = \delta^{\alpha\beta} f(0)^{q-1} \frac{\J}q \delta(t_{12}) \left( \ba{cccc} 0 & -1 \\ 1 & 0  \ea \right), \quad \Sigma_{a}^{\alpha\beta}(\omega) =\delta^{\alpha\beta} f(0)^{q-1} \frac{\J}q \left( \ba{cccc} 0 & -1 \\ 1 & 0  \ea \right),
\eea
where $\hat \Sigma_a(\omega) = \int dt_{12}\hat \Sigma_a(t_{12}) e^{i\omega t_{12}}$. Here the limit $\J T \gg 1$ is implicit, as we are interested in the long-time behaviors of R\'enyi entropy, so the Fourier transform becomes an integral. We also solve the Schwinger-Dyson equation~(\ref{eq:SD1},~\ref{eq:SD2}) numerically in Appendix~\ref{append:finite_T} for finite $T$, and find excellent agreement with the analytic solution we give in the following. 

Plugging the self-energy into~(\ref{eq:SD1}) gives
\bea
	G_{a}^{\alpha\beta}(\omega) &=& \delta^{\alpha\beta} \frac1{\omega^2 +  f(0)^{2q-2} (\frac{\J}{q})^2} \left( \ba{cccc} i \omega & -f(0)^{q-1} \frac{\J}q \\ f(0)^{q-1} \frac{\J}q & -i \omega   \ea \right), \\
	G_{a}^{\alpha\beta}(t_1,t_2)  &=& \delta^{\alpha\beta} \frac{e^{- f(0)^{q-1} \frac{\J}q |t_{12}|}}2 
\left( \ba{cccc} \sgn(t_{12}) & -1 \\ 1 & -\sgn(t_{12})   \ea \right).
\eea
Comparing it to the ansatz, we find $f(t_{12}) = e^{- \frac{\J}q |t_{12}|}$, so the replica diagonal solution is
\bea\label{eq:diagonal}
	G_{a}^{\alpha\beta}(t_1,t_2) = \frac{e^{-\frac{\J}q |t_{12}|}}2 \delta^{\alpha\beta}
\left( \ba{cccc} \sgn(t_{12}) & -1 \\ 1 & -\sgn(t_{12})   \ea \right)
, \quad 	\Sigma_a^{\alpha\beta}(t_1,t_2) = \frac{\J}{q} \delta^{\alpha\beta} \delta(t_{12}) \left( \ba{cccc} 0 & -1 \\ 1 & 0  \ea \right).
\eea

Now we claim that the above function (\ref{eq:diagonal}) is still a solution to the Schwinger-Dyson equation with finite coupling $\V>0$. The reason is two-fold: (1) for the replica-diagonal solution, the $\V$ term in (\ref{eq:SD2}) is only non-vanishing on intra Keldysh contour $s=s'$ term due to the twist operator, and (2) the self-energy only depends on the Green's function at $t_{12}=0$ which is vanishing on the intra Keldysh contour $s=s'$ term. So if one plugs~(\ref{eq:diagonal}) into the Schwinger-Dyson equation, the $\V$ term in~(\ref{eq:SD2}) vanishes and the it solves the equation. 

Besides the replica diagonal solution, the twist operator induces new replica non-diagonal solutions that are the analog of the wormhole solutions found in~\cite{Penington:2019replica}. We assume the subsystem $a=2$ still hosts the replica diagonal solution, $G_{2}^{\alpha\beta}, \Sigma_{2}^{\alpha\beta}  \propto \delta^{\alpha\beta}$, and find that the subsystem $a=1$ supports a replica non-diagonal solution. To get a replica non-diagonal solution, the nontrivial part must come from the twist operator in~(\ref{eq:SD2}), corresponding to the inter Keldysh contour correlation function that crosses the twist operator as shown in Fig.~\ref{fig:contour}. The self-energy of subsystem $a=1$ is given by
\bea\label{eq:self-energy1}
	\Sigma_{1,-+}^{\alpha\beta} = \frac{\delta(t_{12})}q \Big[ \J (2G_{1,-+}^{\alpha\beta})^{q-1} + \V (2G_{1,-+}^{\alpha\beta})^{q/2-1} \sum_{\gamma} \epsilon^{\alpha \gamma}  (2G_{2,-+}^{\gamma\beta})^{q/2} \Big].
\eea
For a diagonal solution $\Sigma_{1,-+}^{\alpha\beta} \propto \delta^{\alpha\beta}$, the second term in~(\ref{eq:self-energy1}) vanishes, and the equation reduces to the diagonal solution~(\ref{eq:diagonal}) as we have shown. However, the second term in~(\ref{eq:self-energy1}) also suggests a replica non-diagonal solution, $\Sigma_{1,-+}^{\alpha\beta} \propto \epsilon^{\alpha\beta}$. 

This leads us to consider a replica non-diagonal ansatz for subsystem $a=1$ (and a replica diagonal ansatz for subsystem $a=2$),
\bea\label{eq:non-diagonal}
	\hat G_1 = \frac{f_1(t_{12})}2  \left( \ba{cccc} \sgn(t_{12}) & - \tilde\epsilon^T \\ \tilde \epsilon & -\sgn(t_{12})   \ea \right), 
	\quad \hat G_2 = \frac{f_2(t_{12})}2  \left( \ba{cccc} \sgn(t_{12}) & - 1 \\ 1 & -\sgn(t_{12})   \ea \right), 
\eea
where the replica indices are implicit, and $\tilde\epsilon^{\alpha\beta} \equiv \sgn(\alpha-\beta) \delta^{\alpha+1,\beta}$. Note that $\tilde \epsilon^T \epsilon = 1$. The sign prefactor in~$\tilde\epsilon$ is due to an emergent time ordering between different replicas when different replicas develop nonvanishing correlations. This is also confirmed by the numerical solutions (see Fig.~\ref{fig:solution}). A detailed derivation is given in Appendix~\ref{append:non-diagonal}, where one finds that $f_1$ and $f_2$ are $f_1(t_{12})= f_2(t_{12}) = e^{- \frac{\J+\V}q |t_{12}|}$. The replica non-diagonal solutions read
\bea\label{eq:non-diagonal2}
	\hat G_1 = \frac{e^{- \frac{\J+\V}q |t_{12}|}}2  \left( \ba{cccc} \sgn(t_{12}) & - \tilde\epsilon^T \\ \tilde \epsilon & -\sgn(t_{12})   \ea \right), 
	\quad \hat G_2 = \frac{e^{- \frac{\J+\V}q |t_{12}|}}2  \left( \ba{cccc} \sgn(t_{12}) & - 1 \\ 1 & -\sgn(t_{12})   \ea \right).
\eea
Thus, we have found both replica diagonal and non-diagonal solutions. We check these solutions by numerically iterating the Schwinger-Dyson equation~(\ref{eq:SD1},~\ref{eq:SD2}). In the numerical calculations, we focus on $n=2, 3$. As shown in Fig.~\ref{fig:solution}, the agreement between the analytics and the numerical solutions is quite good. Note the small contributions near the twist operators in the replica non-diagonal case. The analytical result above corresponds to the limit of large $T$ where these boundary contributions can be neglected, but they do give important contributions to the on-shell action as we discuss below.

\begin{figure}
\centering
\subfigure[]{
	\includegraphics[width=4cm]{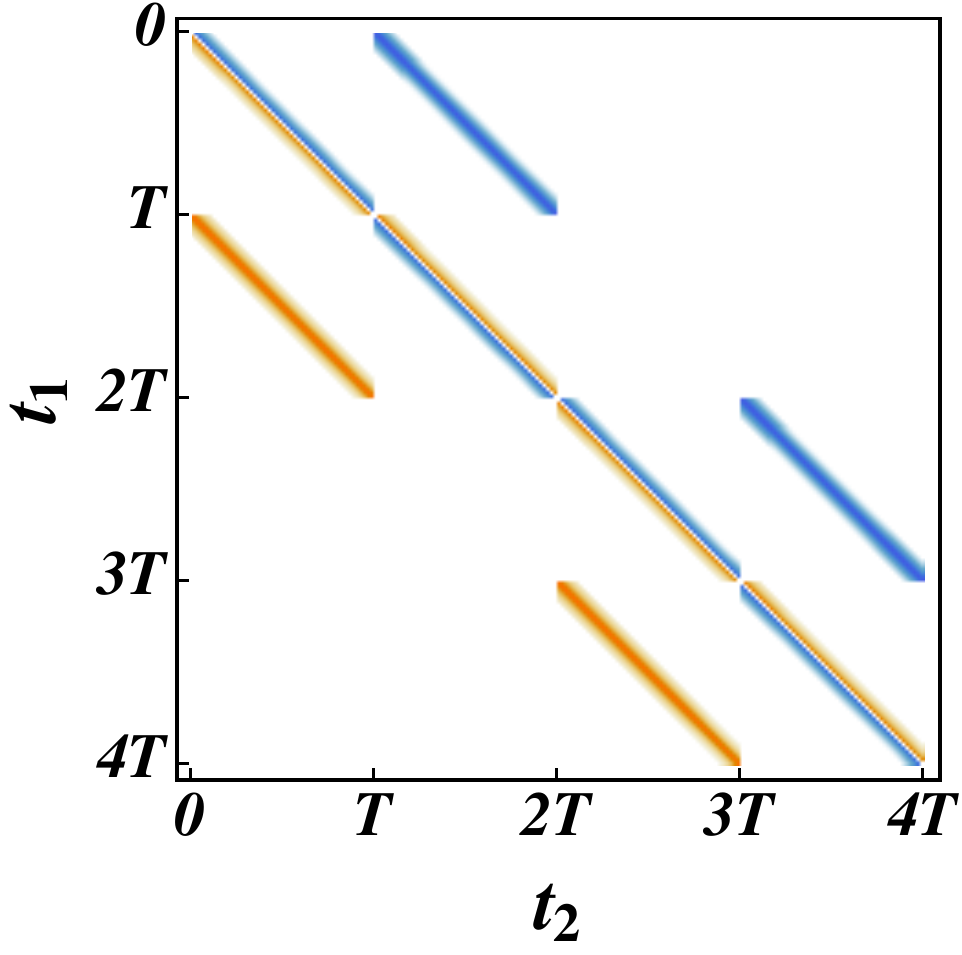}} \qquad \qquad
\subfigure[]{
	\includegraphics[width=4cm]{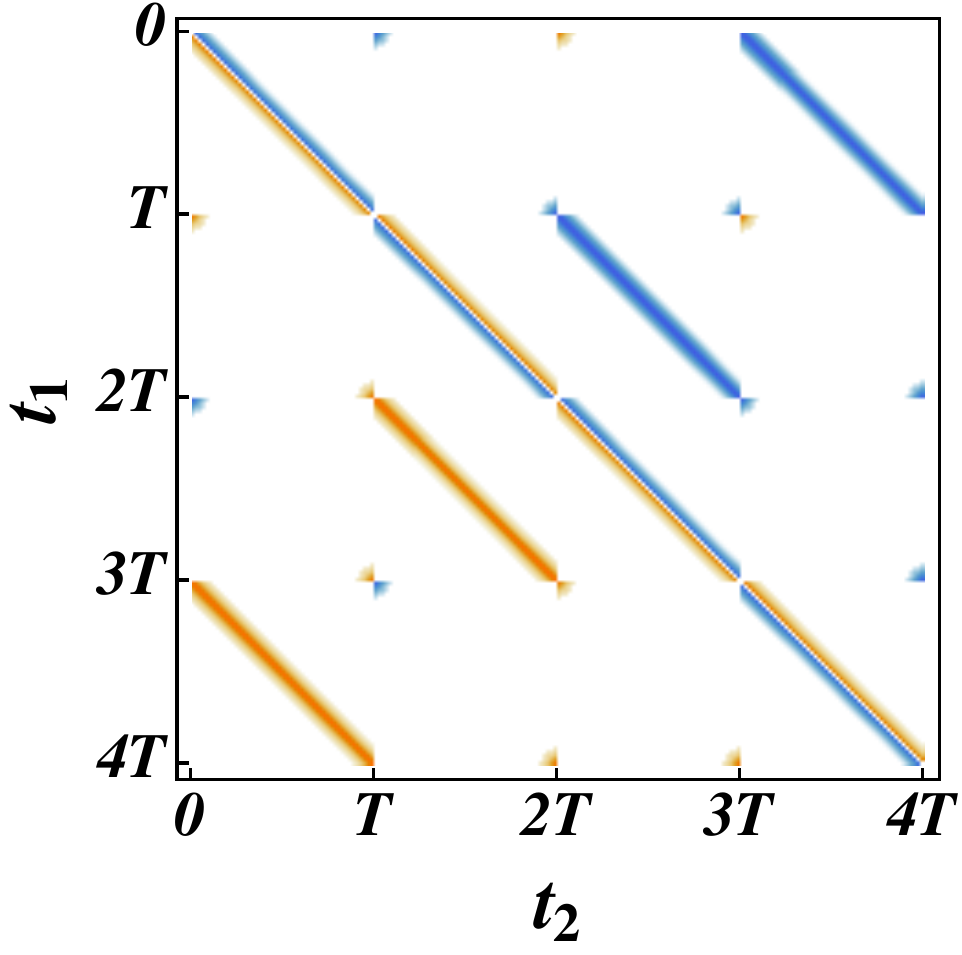}} \\
\subfigure[]{
	\includegraphics[width=4cm]{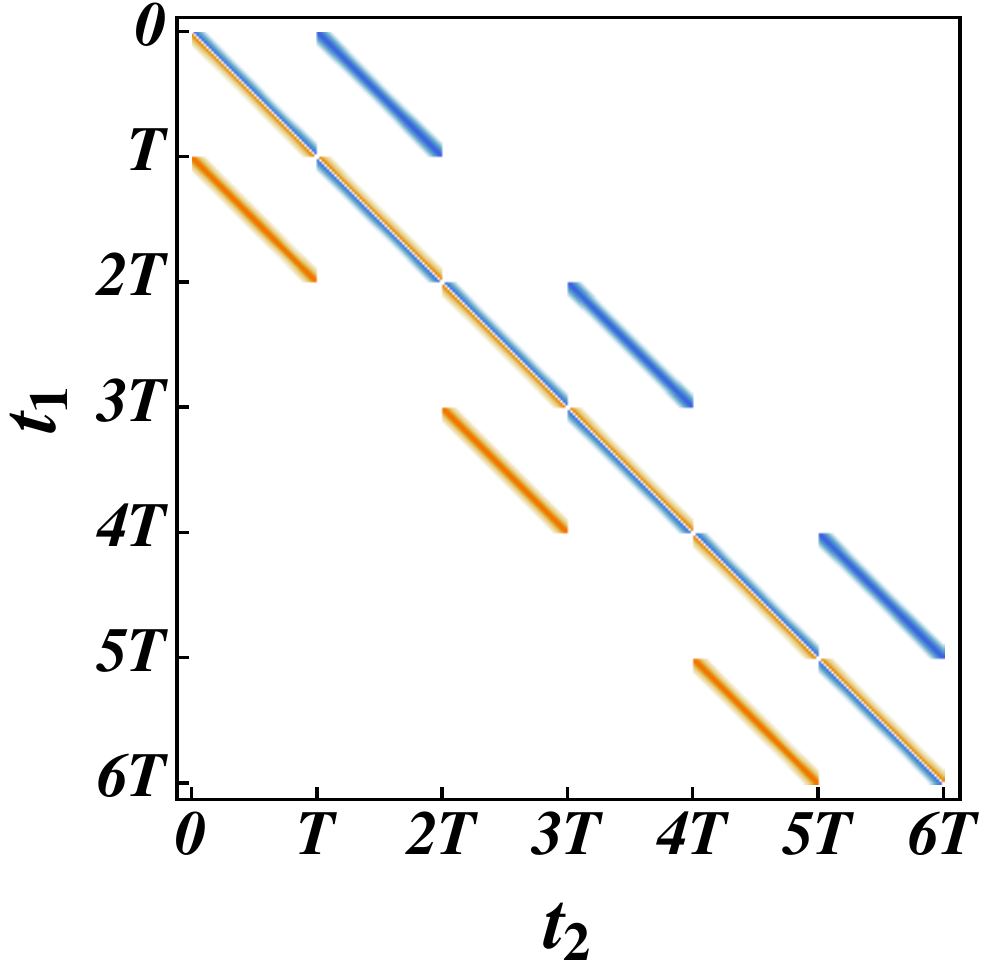}} \qquad \qquad
\subfigure[]{
	\includegraphics[width=4cm]{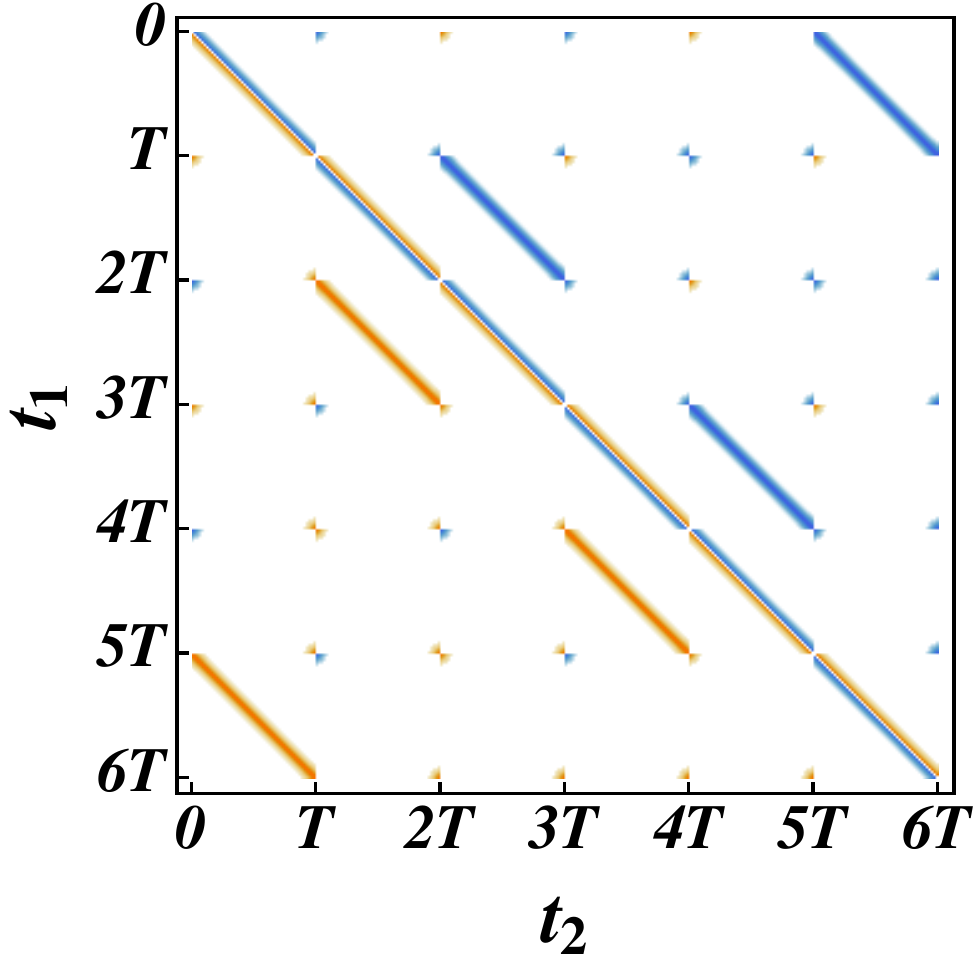}} 
	\caption{The numerical replica diagonal and non-diagonal solutions for $n=2$ (a,b), and for $n=3$ (c,d). The coordinates are arranged such that $(2k T, (2k+2) T)$ belongs to the $k$-th replica, and within each replica, $(2k T, (2k+1)T) $ [$( (2k+1) T, (2k+2)T) $] is the forward $s=+$ (backward $s=-$) contour. Small nonvanishing values appear at the locations of the twist operators. This has important contributions to the onshell action, and we discuss it in Appendix~\ref{append:onshell}. \label{fig:solution}}
\end{figure}

\subsection{Page curve from saddle points}

We now show that the replica diagonal and non-diagonal solutions lead to the linear increase and the saturation of the R\'enyi entropy, respectively. We first give analytic results for R\'enyi entropy from the two saddle point solutions, and then numerically evaluate the onshell action to verify our analytic results.

Recall that the replica diagonal solution~(\ref{eq:diagonal}) is actually the same as the solution of Schwinger-Dyson equation~(\ref{eq:SD2}) when $\V=0$. This means the replica diagonal solution is a solution of replicated action without inserting the twist operator because the twist operator effect is proportional to $\V$. As a result, the first line in~(\ref{eq:action}) counts the total Hilbert space dimension. For each replica system, we have two subsystems $a=1,2$ each hosting $N$ Majorana fermions (remember we have doubled the Hilbert space to prepare the maximally entangled initial state within each subsystems), thus the total Hilbert dimension is $2^{n N}$. In the second line of~(\ref{eq:action}), for replica diagonal solution the inter Kelysh contour component $s\ne s'$ is zero. The intra Keldysh contour component $s=s'$ leads to a linear increase of Renyi-$n$ entropy\footnote{The factor $\delta(x) \sgn(x)^q$ seems to give a vanishing result. But we can consider a smeared out $\delta$-function, which leads to a nonvanishing result. We verify numerically in Appendix~\ref{append:onshell} that this smearing procedure indeed gives the correct linear growth.}, i.e.,
\bea\label{eq:onshell_diagonal}
	- \frac{I^{(1)}}N &=&  n \log 2 - \frac{n \V}{q^2} \int dt_1 dt_2 \delta(t_{12}) \big( e^{-\frac{\J}q |t_{12}|} \sgn(t_{12}) \big)^q =n \log 2 - \frac{n \V T}{q^2} , \\
	\quad \frac{e^{- I^{(1)}}}{Z^n} &=& e^{-\frac{n N \V T}{q^2}},
\eea
where the first term in $I^{(1)}$ is canceled by the denominator which is the Hilbert space of $n$ replicated systems $Z^n = 2^{n N}$.

For the replica non-diagonal solution, we first show that it leads to a time-independent on-shell action\footnote{For finite $T$, we need to consider the effects of the twist operators which serve as boundary conditions at the end of the Keldysh contour. The on-shell action from the replica non-diagonal solution is then not time independent and will receive important corrections. We discuss this correction in Sec.~\ref{sec:finite_time} and in Appendix~\ref{append:twist}.}. The on-shell action is a function of $\J T $ and $\V T$, $I=I[\J T, \V T]$, so its time derivative reads
\bea\label{eq:derivative}
	\frac{d I}{dT} &=& \frac{\J}{T} \frac{\partial I}{\partial \J} + \frac{\V}{T} \frac{\partial I}{\partial \V} \\
	&=& -\frac{N}{T} \int dt_1 dt_2 \delta(t_{12}) c_{ss'} \Big[ \frac{\J }{4 q^2} \sum_a (2G_{a,ss'}^{\alpha\beta})^q + \frac{\V}{2q^2} (2G_{1,ss'}^{\alpha\beta})^{q/2} g^{\alpha\gamma}_s g^{\beta\delta}_{s'} (2G_{2,ss'}^{\gamma\delta})^{q/2} \Big].
\eea
Plugging the replica non-diagonal solution~(\ref{eq:non-diagonal2}) into the time derivative, we have
\bea\label{eq:derivative2}
	\frac{d I^{(2)}}{dT} &=& \frac{N}{T} \int dt_1 dt_2 \delta(t_{12}) (1+1 - 1 -1) \Big[ \frac{\J +\V}{2 q^2} \big( e^{-\frac{\J+\V}q |t_{12}|} \big)^q \Big] = 0 ,
\eea
where the vanishing prefactor is coming from summing over $c_{ss'}$. The difference between replica diagonal and non-diagonal solutions originates from the twist operator, and the non-diagonal solution has a nontrivial contribution from the inter Keldysh contour component, so the time derivative vanishes. Because the bulk of the on-shell action vanishes, its value is determined by boundary effects near the twist operators.

Explicitly, the onshell action for the replica non-diagonal solution is given by
\bea \label{eq:action_non-diagonal}
	&& -\frac{I^{(2)}}{N} = \sum_a \log \Pf [ \hat \sigma^z \partial_t - \hat \Sigma_a] \nn\\
	 && + (1-q) \int \delta(t_{12})  c_{ss'} \Big[ \sum_a \frac{\J}{4q^2}  (2G_{a,ss'}^{\alpha\beta})^q  +  \frac{\V}{2q^2} (2G_{1,ss'}^{\alpha\beta})^{q/2} g^{\alpha\gamma}_s g^{\beta\delta}_{s'} (2G_{2,ss'}^{\gamma\delta})^{q/2} \Big], \\
	 && \qquad \quad = \sum_a \log \Pf [ \hat G_a^{-1}],
\eea
where the second line vanishes for the same reason as in~(\ref{eq:derivative2}), and we use~(\ref{eq:SD1}) to get the third line. Hence, the on-shell action is determined by the Pfaffian of the Green's function.

It is convenient to approach the calculation of the Pfaffian by recalling that the $a=2$ subsystem still hosts the replica diagonal solution, which is also a solution when there is no twist operator in~(\ref{eq:action}). Consider the action without twist operators,
\bea \label{eq:action_no_twist}
	- \frac{I^{(0)}}{N} &=& \sum_a \Big[ \log \Pf [ \partial_t \hat \sigma^z - \hat \Sigma_a] + \int dt_1 dt_2  \big( - \frac{1}2  \Sigma_{a,ss'}^{\alpha\beta} G_{a,ss'}^{\alpha\beta} + \frac{\J }{4 q^2} \delta(t_{12}) c_{ss'} (2G_{a,ss'}^{\alpha\beta}(t_1,t_2))^q \big) \Big] \nn\\
	&& + \frac{\V}{2q^2} \int dt_1 dt_2 \delta(t_{12}) c_{ss'}(2G_{1,ss'}^{\alpha\beta}(t_1,t_2))^{q/2} (2G_{2,ss'}^{\alpha\beta}(t_1,t_2))^{q/2},
\eea
where the replica indices are trivial because there are no twist operators. Similar to the previous calculation, one can show that the large-$N$ solution of~(\ref{eq:action_no_twist}) for both subsystems is the same as the diagonal solution in subsystem $a=2$ given in~(\ref{eq:non-diagonal}). We copy it here using the same symbol for convenience,
\bea
	\hat G_2 = \frac{e^{- \frac{\J+\V}q |t_{12}|}}2  \left( \ba{cccc} \sgn(t_{12}) & - 1 \\ 1 & -\sgn(t_{12})   \ea \right).
\eea
The onshell action of~(\ref{eq:action_no_twist}) is simply $- \frac{I^{(0)}}{N} = 2 \log \Pf [ \hat G_2^{-1}]$ which is actually the logarithm of Hilbert space dimension, i.e., $e^{-I^{(0)}} = Z^n$. 
Using this action as denominator, the R\'enyi entropy from replica non-diagonal solution can be written as
\bea \label{eq:onshell_non-diagonal}
	\frac{e^{-I^{(2)}}}{Z^n} = e^{-(I^{(2)}-I^{(0)})} =\exp N \Big( \log \Pf [ \hat G_1^{-1}] - \log \Pf [ \hat G_2^{-1}] \Big),
\eea
where $\hat G_1$ and $\hat G_2$ are given by~(\ref{eq:non-diagonal2}). However, as discussed, (\ref{eq:non-diagonal2}) neglects the effect of twist operators near the boundary, which must be included to get the correct answer. Thus we need to calculate the Pfaffian using the full numerical solution where the effect of twist operators is automatically included. We evaluate the Pfaffian for $n=2,3,...,10$ in Appendix~\ref{append:onshell} and find that
\bea\label{eq:page}
	 \log \Pf [ \hat G_1^{-1}] - \log \Pf [ \hat G_2^{-1}]  = (1-n) \log 2.
\eea
We expect that this result holds true for any integer $n$ because it gives the correct answer for Brownian evolutions where the R\'enyi entropy is maximized at late time.

\begin{figure}
	\centering
\subfigure[]{
	\includegraphics[width=6cm]{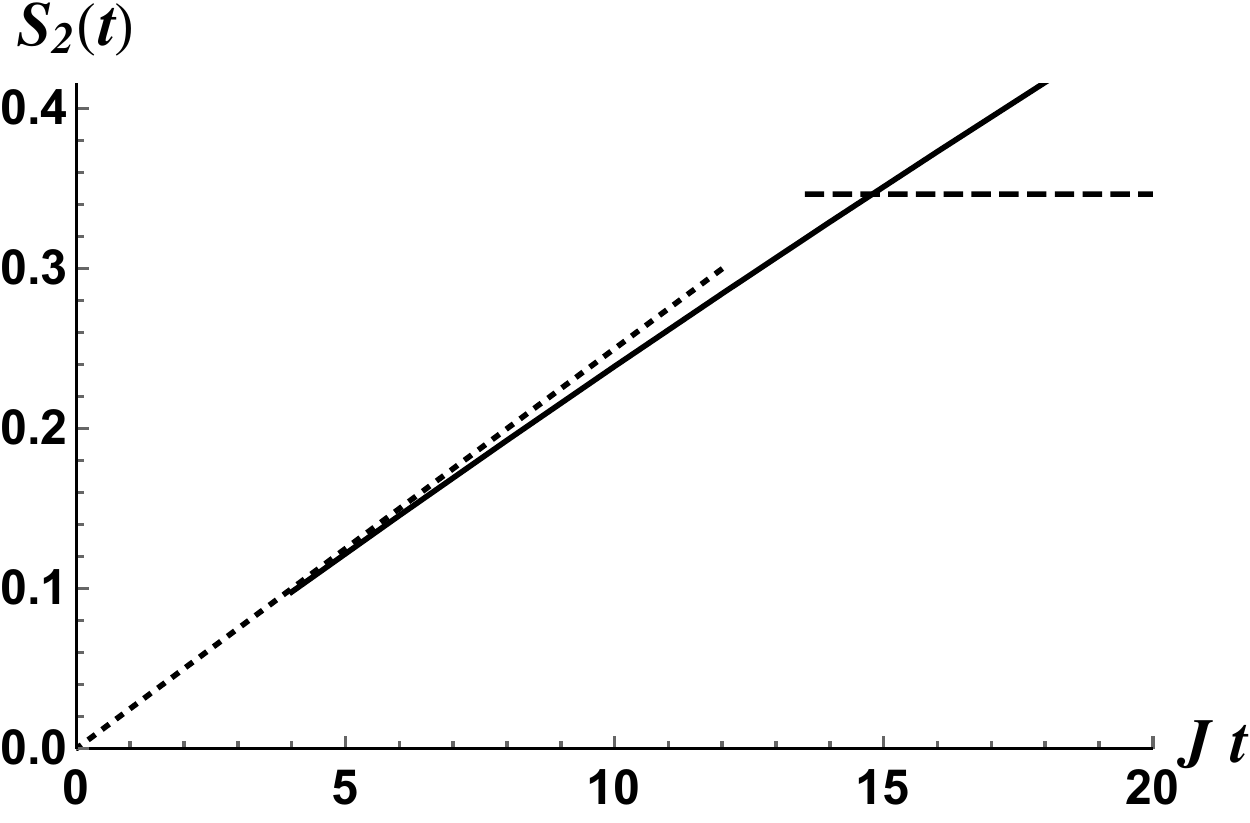}}\qquad
\subfigure[]{
	\includegraphics[width=6cm]{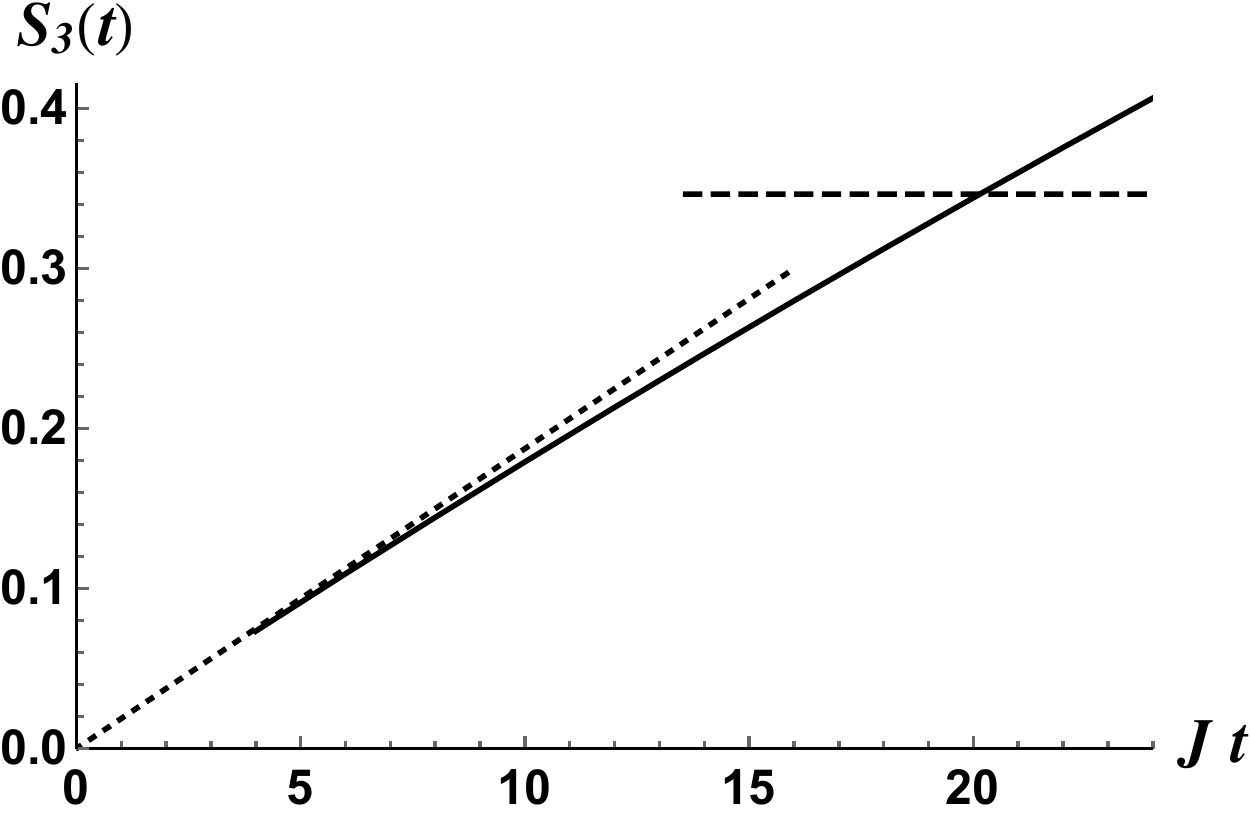}}
\caption{(a) The second R\'enyi entropy and (b) the third R\'enyi entropy per Majorana fermion. The solid line is from the replica diagonal solution, which grows linearly. The dashed line is from the replica non-diagonal solution that takes over after Page time. The value of the dashed line is $\frac12 \log 2$. The dotted line plots the first line in~(\ref{eq:saddle_result}) for comparison. \label{fig:page_curve} }
\end{figure}

From the on-shell actions of the replica diagonal solution (\ref{eq:onshell_diagonal}) and replica non-diagonal solution (\ref{eq:onshell_non-diagonal}), we find that the $n$-th R\'enyi entropy at time $T$ is
\bea \label{eq:Renyi_saddle}
	e^{-(n-1) S_n(T)} &=&  \frac{e^{-I^{(1)}}+ e^{-I^{(2)}}}{Z^n} = e^{-\frac{ n N \V T}{q^2}} + 2^{-N (n-1) }, \\
	S_n(T) &=&  \frac1{1-n} \log\Big[ e^{- \frac{n N \V T}{q^2}} + 2^{-N (n-1)} \Big]  \rightarrow 
	\begin{cases}
		\frac{n}{n-1} \frac{N \V T}{q^2}, \quad & T \ll T^\ast \\
		N \log2 , \quad & T \gg T^\ast
	\end{cases}
\eea
which is the Page curve in the coupled Brownian SYK models and $T^\ast = \frac{n-1}n \frac{q^2}{\V} \log2 + \O(1/N)$. To convert the result to entropy per Majorana, we notice that the number of Majorana fermion in the doubled Hilbert space of subsystem $a=1$ is $2N$, so we have
\bea\label{eq:saddle_result}
	\S_n(t) &\equiv& \frac{S_n(T=2t)}{2N} = 
	\begin{cases}
		\frac{n}{n-1} \frac{\V}{q^2} t, \quad & t \ll t^\ast \\
		\frac{1}2 \log2 , \quad & t \gg t^\ast
	\end{cases}
\eea
where $t^*=T^*/2$ is the Page time. It is interesting to note the Page time increases for increasing $n$, but remains finite for $n \rightarrow \infty$. We also evaluate the R\'enyi entropy of the replica diagonal and non-diagonal solution numerically. The results for $n=2,3$ are shown in Fig.~\ref{fig:page_curve}.

So far we have mainly discussed the R\'enyi entropy. It is an interesting question to understand the entanglement entropy which can be obtained in principle by analytically continue $n \rightarrow 1$ of the R\'enyi entropy. There are two ways of doing the analytical continuation depending on when to take the limit. One can keep the action off-shell and take the limit near $n \approx 1$ to get the corresponding saddle-point equation~\cite{Almheiri:2019replica, Penington:2019replica}. In our case, since $n$ is the dimension of various matrices $G, \Sigma, g$, etc, it is not clear how to do it. Another way is to take the analytically continuation of the on-shell saddle-point action. This amounts to evaluate all possible saddle-point solutions at general $n$. We have evaluated the fully connected replica non-diagonal solution which is responsible for the saturation of the R\'enyi entropy at long times. It is reasonable to expect that there are lower symmetric saddle-point solutions that also contribute to the action. This is also the reason that a naive $n \rightarrow 1$ limit of~(\ref{eq:Renyi_saddle}) is ill-defined. For $n=3$ case, we explicitly evaluate different replica non-diagonal saddle-point solutions (not shown in the paper). We hope to generalize the calculation to arbitrary $n$, and then analytically continue the result to $n \rightarrow 1$ in the future work.

\subsection{Finite time effects} \label{sec:finite_time}

So far we showed that the replica diagonal solution gives rise to linear entropy growth while the replica non-diagonal solution gives rise to entropy saturation at long times. When the non-diagonal saddle dominates, there were small contributions localized near the twist operator that lead to important effects. Here we discuss the effects of these contributions on the timescale for entropy saturation.

In fact, there are two important times in the problem. The first is the time at which the two saddles exchange dominance. We refer to this as the Page time and note that, at large $N$, it is independent of $N$. The second is the time for the entropy to reach within a few bits of its maximal value. We refer to this time as the strong scrambling time. As we now show, this time scales like $\log N$ at large $N$.

The key point is that in the replica non-diagonal saddle, boundary effects contribute a term in the on-shell action of the form
\begin{equation}
    (1-n) N b_n(T) e^{- \lambda T}
\end{equation}
where $b_n$ grows no faster than polynomial in $T$, and $\lambda$ is a local interaction scale, which in our case is given by $\lambda =2( \J + \V)/q$ at large $q$ limit. After the Page time, the Renyi entropy is thus
\begin{equation}
   S_n =  N \log 2 - N b_n(T) e^{-\lambda T}.
\end{equation}
Hence, as $N$ goes to infinity at fixed $T$, the entropy differs from its saturation value by an amount extensive in $N$. On the other hand, by taking $T \sim \frac1{\lambda}\log N$, the correction term can be made order unity instead of order $N$. At these times, the entropy is therefore within a few bits of its saturation value. 
This form of the on-shell action follows from the exponential decay of correlations. In the limit of large time, the twist operator contribution is local and cannot depend on the temporal extent of the system. Since correlations are exponentially decaying in time, it follows that finite time effects must vanish exponentially fast, up to a polynomial prefactor. A detailed analysis of this physics is possible in the large $q$ limit as discussed in Appendix~\ref{append:twist}.

\section{Purity from operator dynamics} \label{sec:master}

Here we study the entropy dynamics from the perspective of operator dynamics, focusing on the purity, $e^{-S_2}$. This approach is complementary to the path integral approach, including offering easier access to finite $N$ corrections.

\subsection{Operator dynamics of the coupled Brownian SYK models}

As before, we consider a Hilbert space made up of $N_1+N_2$ Majorana fermions. 
The Majorana fermions satisfy $\{\psi_{i,a}, \psi_{j,b} \} = \delta_{ab}\delta_{ij}$, and form an orthonormal basis 
\bea\label{eq:basis}
	\Gamma_{A,B} = i^{[(|A|+|B|)/2]} 2^{(|A|+|B|)/2} [\psi_1]_A [\psi_2]_B, \quad [\psi_a]_A=\psi_{j_1,a}...\psi_{j_{|A|},a},
\eea 
where $A=j_1...j_{|A|}$ is an ascending list of length $|A|$ and $[x]$ is the largest integer less than or equal to $x$. In particular, we use $\Gamma_{A,0} = \Gamma_{A,\emptyset}$ and $\Gamma_{0,A} = \Gamma_{\emptyset,A}$ to denote the basis locating solely in subsystem $a=1,2$. 

We again take $N_1 = N_2$ for simplicity, so the Hilbert space dimension is $2^N$. 
The inner product of operators is $\Tr[\O^\dag \O']$ for any two operators $\O$ and $\O'$. 
We can decompose any operator by the basis~(\ref{eq:basis})
\bea
	\O = \sum_{A,B} c_{A,B} \Gamma_{A,B}, \quad c_{A,B} = \frac1{2^N}\Tr[\O \Gamma_{A,B}].
\eea
If the operator is normalized according to $2^{-N} \Tr[\O^\dag \O]=1$, implying $\sum_{A,B} |c_{A,B}|^2 = 1$, then because unitary evolution preserves the normalization, we can interpret $|c_{A,B}|^2$ as the probability of finding the operator $\O$ in basis operator $\Gamma_{A,B}$. Since the disorder averaged theory has an emergent $SO(N) \times SO(N)$ symmetry for each of the subsystem, one expects the dynamics depends only on the length of operators not on the specific list $A$. So we define the probability distribution of an operator in subsystem $a=1,2$ with length $m,m'$ to be
\bea\label{eq:pmm}
	&& p_{m,m'}(t) = \sum_{|A|=m,|B|=m'} |c_{A,B}(t)|^2 =\sum_{|A|=m,|B|=m'} 2^{-2N} |\Tr[\O(t) \Gamma_{A,B}]|^2. 
\eea
We are interested in finding the master equation governing the time development of this probability distribution. 

In terms of the basis, the Hamiltonian~(\ref{eq:Hamiltonian}) can be rewritten as
\bea \label{eq:H}
	H = \sum_{A, B} J_{A,B}  \Gamma_{A,B}^{(q)}, \quad
	J_{A,0} \equiv 2^{-q/2} J_{A}^1, \quad J_{0,A} \equiv 2^{-q/2} J_{A}^2, \quad J_{A,B} \equiv 2^{-q/2} V_{A,B},
\eea
where the superscript $\Gamma^{(q)}$ implies the length of basis is $q$. The short-hand summation denotes the same summation as in~(\ref{eq:Hamiltonian}). Due to the Brownian nature of the interactions, we can view the Hamiltonian as a random circuit. At each time step, the evolution operator is generated by $U(d t) = e^{-i H(t) dt}$. When we discretize the time interval by a tiny time step $dt \ll T$, it is easy to check the average over Brownian variables takes the following rule
\bea \label{eq:disorder_average}
	\overline{J_{A,B}(t) dt J_{A',B'}(t') dt} =\delta_{AA'} \delta_{BB'} \delta_{tt'}\sigma_{A,B} dt, \\
	\sigma_{A,0} = \sigma_{0,A} = \frac{q! \J}{2q^2N^{q-1}} \equiv \sigma_0 , \quad \sigma_{A, B} = \frac{(q/2)!^2 \V}{q^2 N^{q-1}} \equiv \sigma_1.
\eea
In the following, the over line denoting the disorder average is omitted for notational simplicity.

For a generic operator $\O$, the infinitesimal evolution $\O(t+dt) = U^\dag(dt) \O(t) U(dt) $ is
\bea
	&& \O(t+dt) = e^{i H dt} \O(t) e^{-i H dt}\\
	&=& \O(t)  + i[H, \O(t)] dt - \frac12 \{ H^2 dt^2, \O(t)\} + H \O(t) H dt^2\\
	&=& \left(1- ( 2  C_N^q  \sigma_{0} +   (C_N^{q/2})^2 \sigma_{1}  )dt \right)\O(t)  + i[H, \O(t) ] dt  + \sum_{A,B} \sigma_{A,B} dt \Gamma_{A,B}^{(q)} \O(t) \Gamma_{A,B}^{(q)}.
\eea
where $C_n^m \equiv \frac{n!}{m!(n-m)!}$ denotes the number of $m$-combinations of set with $n$ elements. In the second line, we expand the exponential function in the first line and keep up to the second order in $dt$. In the third line we have performed the disorder average of the $dt^2$ terms by using~(\ref{eq:disorder_average}), and due to the Ito calculus, they become linear in $dt$. Assuming the operator $\O$ is Hermitian, the distribution at time $t+dt$ via the definition~(\ref{eq:pmm}) is given by
\bea\label{eq:pmmt}
	 p_{m,m'}(t+dt) &=& 2^{-2N}\sum_{|A|=m,|B|=m'} |\Tr[O(t+dt) \Gamma_{A,B}]|^2 \\
\label{eq:pmm1}  & =& \left( 1- 2( 2C_N^q \sigma_{0} + (C_N^{q/2})^2 \sigma_{1} ) \right) p_{m,m'}(t) \nn\\
  && + 2^{-2N} \sum_{|A|=m,|B|=m'} \Big( 2\Tr[\O(t) \Gamma_{A,B}] \sum_{C,D} \sigma_{C,D} \Tr[\O(t) \Gamma_{C,D}^{(q)} \Gamma_{A, B} \Gamma_{C,D}^{(q)}]  \nn\\
  && -  \sum_{C,D} \sigma_{C,D}  \Tr^2 (\O(t) [\Gamma_{A,B} ,\Gamma_{C,D}^{(q)} ]) \Big) dt,
\eea
where the summation over $C, D$ is the same as in~(\ref{eq:H}). We have performed disorder average in the second line. 

The master equation of $p_{m,m'}(t)$ can be derived straightforwardly. We leave the detailed derivation in Appendix~\ref{append:master}, and the result is
\bea\label{eq:master}
	\frac{d p_{m,m'}(t)}{dt} &=& -4 \Big[ \sigma_{0} \sum_{k=1, \text{odd}}^{\min(q,m)} C_{N-m}^{q-k} C_{m}^k + \sigma_{0} \sum_{k'=1, \text{odd}}^{\min(q,m')} C_{N-m'}^{q-k'} C_{m'}^{k'}  \nn\\
	&& +  \sigma_{1} \sum_{k=0}^{\min(q/2,m)} \sum_{k'=0}^{\min(q/2,m')} \frac{1-(-1)^{k+k'}}{2} C_{N-m}^{q/2-k} C_{m}^k  C_{N-m'}^{q/2-k'} C_{m'}^{k'} \Big] p_{m,m'}(t), \nn\\
	&& + 4 \Big[ \sigma_0 \sum_{k=1, \text{odd}}^{\min(q,m)}  C_{N-(m+q-2k)}^k C_{m+q-2k}^{m-k} p_{m+q-2k,m'}(t) \nn\\
	&& + \sigma_0 \sum_{k'=1, \text{odd}}^{\min(q,m')}   C_{N-(m'+q-2k')}^{k'} C_{m'+q-2k'}^{m'-k'}  p_{m,m'+q-2k'}(t)  \nn\\
	&& + \sigma_1 \sum_{k=0}^{\min(q/2,m)} \sum_{k'=0}^{\min(q/2,m')} \frac{1-(-1)^{k+k'}}2 C_{N-(m+q/2-2k)}^k   C_{m+q/2-2k}^{m-k}  \nn\\
	&& \times C_{N-(m'+q/2-2k')}^{k'} C_{m'+q/2-2k'}^{m'-k'}   p_{m+q/2-2k,m'+q/2-2k'}(t) \Big],
\eea
where the first two lines is the out-going rate and the rest is the in-coming rate. It is straightforward but tedious to show that the following distribution is a stationary solution to the master equation
\bea
	p^{(\text{st})}_{m,m'} = 2^{-2N} C_{N}^m C_N^{m'},
\eea
which means the probability in any basis $\Gamma_{A,B}$ is the same, i.e., $|c_{A,B}^{(\text{st})}|^2 = 2^{-2N}$. This is consistent with the expectation of approaching an infinite temperature state in the Brownian evolution, where no any specific basis is preferred.

It is instructive also to check the symmetry of the master equation. First, after the disorder average the model has an $SO(N) \times SO(N)$ symmetry, and this is the reason that the master equation can be reduced $p_{A,B} \rightarrow p_{m,m'}$ depending only on the length of the basis.

Second, depending on the parity of $q/2$, i.e., the interactions between two Brownian SYK models, the model has $Z_2^f \times Z_2^f$ for even $q/2$, i.e., the Fermi parity is conserved separately in two subsystems, or $Z_2^f$ for odd $q/2$, i.e., only the total Fermi parity is conserved. This leads to the result that $p_{m,m'}$ couples only to $p_{m+ q/2 -2k, m' + q/2 - 2k'}$ in~(\ref{eq:master}). If $q/2$ is even, then the master equations for distribution $p_{m,m'}$, $(m, m' ) \in (\text{even}, \text{even})$,  $(\text{even}, \text{odd})$,  $(\text{odd}, \text{even})$, $(\text{odd}, \text{odd})$ decouple. If $q/2$ is odd, then the master equations for distribution $p_{m,m'}$, $m+m' \in \text{even}, \text{odd}$ decouple.

Finally, some special operators are conserved due to the Fermi parity symmetry. For even $q/2$, four operators $\Gamma_{0,0}$, $\Gamma_{\{12...N\},0}$, $\Gamma_{0,\{12...N\}}$, $\Gamma_{\{12...N\},\{12...N\}}$ are conserved, while for odd $q/2$, only two operators $\Gamma_{0,0}$, $\Gamma_{\{12...N\},\{12...N\}}$ are conserved.

\subsection{Purity evolution of a pure state}

Now we relate the purity evolution to the operator dynamics. The discussion in the following is general for any initial density matrix $\rho^\text{(in)}$ in the Hilbert space span by $2N$ Majorana operators. 
The density matrix evolves
\bea
	\rho(t) &=& U(t) \rho^\text{(in)} U^\dag(t) \equiv \sum_{A,B} c_{A,B }(t) \Gamma_{A,B}, \quad  c_{A,B }(0)= c^{(\text{in})}_{A,B},
\eea
so the reduced density matrix is
\bea
	\rho_\text{1}(t)  &=& \Tr_2[ \sum_{A,B} c_{A,B}(t) \Gamma_{A,B} ] =  2^{N/2} \sum_{A} c_{A,0}(t) \Gamma_{A}, \\
	e^{-S_2(t)} &=& \Tr_1[\rho_1(t)^2] = 2^{3N/2} \sum_{A} |c_{A,0}(t)|^2.
\eea
Here $\Gamma_A \equiv  i^{[|A|/2]} 2^{|A|/2} [\psi_1]_A$ is the basis in subsystem $a=1$. In the second equation, we have used the orthonormal property of the basis $\Gamma_A$, namely, $\Tr_1 [\Gamma_A \Gamma_B] = 2^{N/2} \delta_{AB} $.

The evolution of the operator wavefunction $c_{A,B}$ is captured by the master equation~(\ref{eq:master}), so the purity dynamics is also dictated by the master equation. 
We should notice that the density matrix is not normalized with respect to $2^{-N} \Tr[\O^\dag \O] = 1$.
For simplicity, let us consider pure initial state so that $\Tr[ \rho^2] =1 $.
We can look at normalized operator,
\bea
	\O_\rho = 2^{N/2} \rho = \sum_{A,B} 2^{N/2} c_{A,B}(t) \Gamma_{A,B}, \quad p_{m,m'}^\O =  \sum_{|A|=m, |B|=m'} 2^N |c_{A,B}|^2.
\eea
So the purity is
\bea
	e^{-S_2(t)} &=& 2^{N/2} \sum_m p_{m,0}^\O(t),
\eea
which is $2^{N/2}$ times the probability of finding the operator in subsystem $a=1$.

\subsection{Setup}

We consider the following state (not be confused with the TFD state considered in the previous section), $|\infty \rangle $, that is similar to the Kourkoulou-Maldacena state~\cite{Kourkoulou:2017pure}
\bea
	&& (\psi_{2j-1,a} + i \psi_{2j,a} )|\infty \rangle = 0, \quad \langle \infty | (\psi_{2j-1,a} - i \psi_{2j,a} ) = 0, \quad \forall a=1,2, \quad \forall j = 1,...,N.
\eea
So the initial density matrix is
\bea \label{eq:initial}
 	\rho^{(\text{in})} &=& |\infty\rangle \langle \infty | = \prod_{j=1}^{N/2} (\frac12 - i \psi_{2j-1,1} \psi_{2j,1}) \prod_{j=1}^{N/2}(\frac12 - i\psi_{2j-1,2} \psi_{j,2}) \\
	&=& 2^{-N} \prod_{j=1}^{N/2} (1- \Gamma_{\{2j-1,2j\},0}) (1- \Gamma_{0,\{2j-1,2j\}}) \equiv \sum_{A,B} c^{(\text{in})}_{A,B} \Gamma_{A,B},
\eea
where in the second line we used the basis defined in~(\ref{eq:basis}). The probability distribution of the normalized density matrix operator at time zero is 
\bea \label{eq:initial_distribution}
	p_{2m,2m'}^\O(0) = 2^{-N} C_{N/2}^m C_{N/2}^{m'},
\eea
which implies the purity is one (equivalently, the second R\'enyi entropy is zero),
\bea
	e^{-S_2(0)} =2^{N/2} \sum_m p_{m,0}^\O(0) = 2^{-N/2} \sum_{m=0}^{N/2} C_{N/2}^m = 1.
\eea
This is consistent with the fact that the initial state is a product state between the two subsystems.

\subsection{Page curve from the master equation}

The system is prepared in a pure state with the initial distribution given by~(\ref{eq:initial_distribution}). Though it is not easy to solve the master equation exactly, we can obtain the final probability distribution. Considering the symmetry of the master equation, the final probability distribution depends on the parity of $q/2$. If $q/2$ is even, the final distribution is
\bea
	p_{m, m'} = \begin{cases} \frac{1-2^{2-N}}{2^{2N-2}-4} C_N^{m} C_N^{m'}, & \{ m,m' \in 2Z \}  \cap \{ m, m' \ne 0 , N \} \\
				2^{-N}, & (m,m') = (0,0), (0,N), (N,0), (N,N) \\
				0, & m,m' \notin 2Z
	\end{cases}.
\eea
which leads to the purity
\bea
	e^{-S_2} = \frac{2^{N/2}}{2^{N-2} + 1}, \qquad S_2 = \frac{N}2 \log 2 - \log 4 + O(2^{-N}).
\eea
The deficit of $\log 4$ is due to $Z_2^f \times Z_2^f$ symmetry, since the Hilbert space has four decoupled sectors. Because the initial state~(\ref{eq:initial}) is prepared in the (even, even) sector, its maximal entropy is given by $\frac{N}2 \log 2 - \log 4$.

On the other hand, if $q/2$ is odd, the final distribution is
\bea
	p_{m, m'} = \begin{cases}  \frac{1-2^{1-N}}{2^{2N-1}-2} C_N^{m} C_N^{m'}, & \{ m+m' \in 2Z\}  \cap \{m, m' \ne 0 , N\} \\
				2^{-N}, & (m,m') = (0,0),  (N,N) \\
				0, & m+m' \notin 2Z
	\end{cases}.
\eea
which leads to the purity 
\bea
	e^{- S_2} = \frac{2^{N/2}}{2^{N-1} + 1}, \qquad S_2 = \frac{N}2 \log 2 - \log 2 + O(2^{-N}).
\eea
The shortage of $\log 2$ is due to the $Z_2^f $ symmetry. The Hilbert space has two decoupled sectors, leading to a deficit of $-\log 2$.

\begin{figure}
	\centering
\subfigure[]{\label{fig:purity}
	\includegraphics[width=4.8cm]{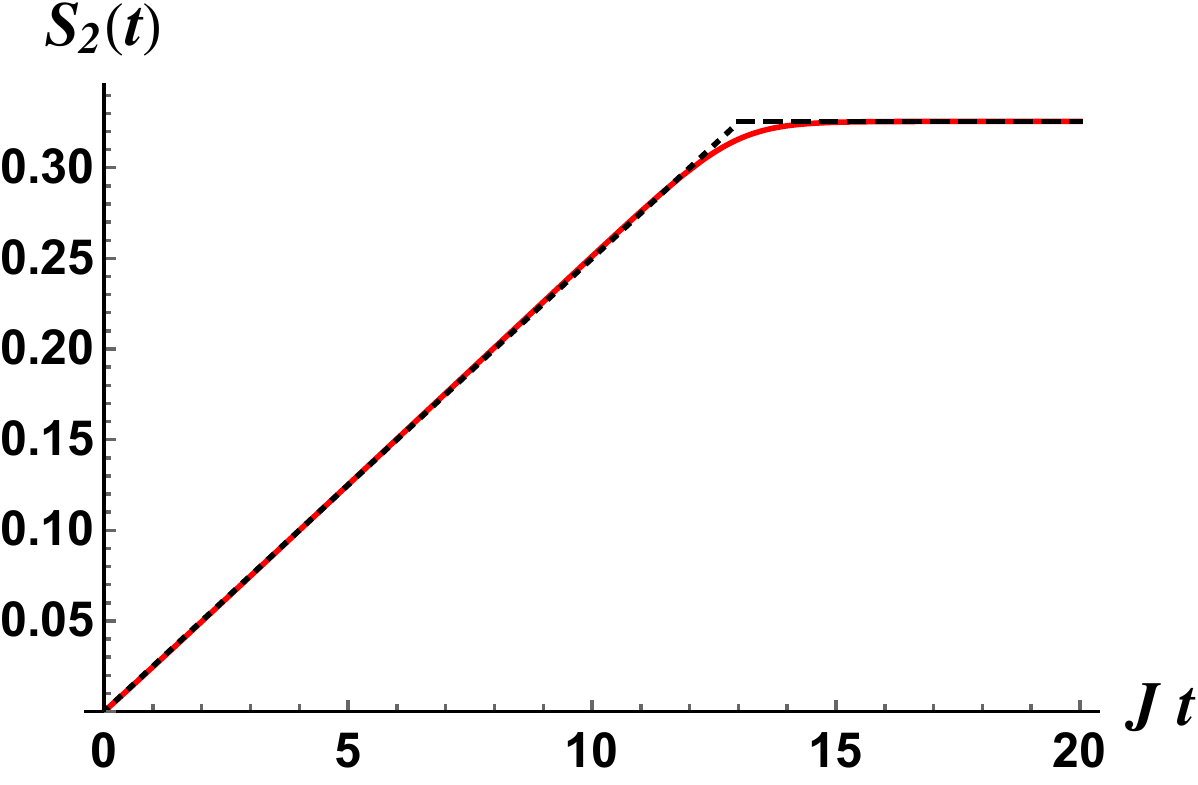}} \qquad
\subfigure[]{\label{fig:exponential}
    \includegraphics[width=5cm]{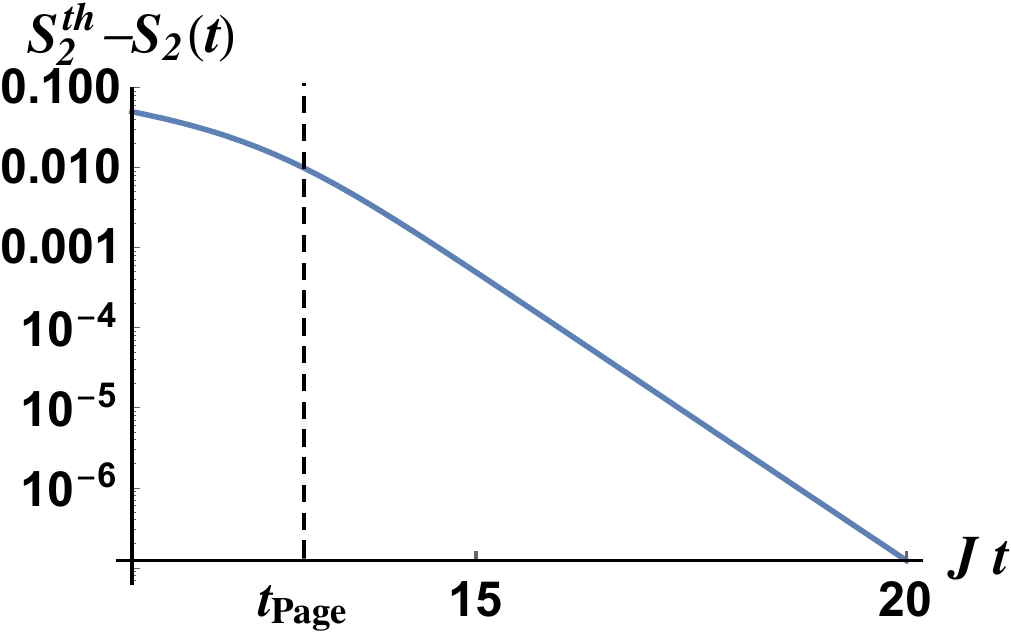}}
	\caption{(a) The Page curve of the second R\'enyi entropy per Majorana fermion $\S_2$ from the master equation. We start from the initial distribution given in~(\ref{eq:initial_distribution}). We choose $N=66$, $q=4$, and $\V/\J=0.2$. The dotted line is the first line in~(\ref{eq:master_result}), and the dashed line is $S_2^{th}= \frac12 \log2 - \frac1{N} \log 4$ where the finite $N$ correction is included. (b) A log linear plot of $S^{th}_2 - S_2(t)$ after the Page time. It indicates an exponential $e^{- \lambda t} $ behavior, consistent with the $\log N$ scrambling time discussed in Sec.~\ref{sec:finite_time}.}
\end{figure}

Thus we expect that under the Hamiltonian~(\ref{eq:H}) evolution, the second R\'enyi entropy will increase from zero to almost the largest value $\frac{N}2 \log 2$. We can get the increase rate, which is the outgoing rate from $P(t)=2^{N/2}\sum_m p_{m,0}(t)$, i.e., the second line in~(\ref{eq:master}). The initial outgoing rate is
\bea
	\frac{d P(t)}{dt} \Big|_{t=0} &=& 2^{N/2} \sum_{m=0}^N \frac{dp_{m,0}}{dt}\Big|_{t=0} = - 2^{N/2} \sum_{m=0}^N 4  \sigma_{1} C_{N}^{q/2} \sum_{k=1, \text{odd}}^{\min(q/2,m)} C_{N-m}^{q/2-k} C_{m}^k p_{m,0}(0), \\
	& \approx & -4 \sigma_{1} C_N^{q/2} \int_0^1 ds \sum_{k=1, \text{odd}}^{q/2} \frac{(1-s)^{q/2-k} s^k}{(q/2-k)! k!} \sqrt{\frac{N}{\pi}} e^{-N (s-1/2)^2}  \\
	& \approx & -4 \sigma_{1} C_N^{q/2} N^{q/2} \frac1{2(q/2)!} \int_0^1 ds [1-(1-2s)^{q/2}] \sqrt{\frac{N}{\pi}} e^{-N (s-1/2)^2} \\ 
	& \approx & - \frac{2N \V}{q^2}.
\eea
where in the second line we take the large-$N$ limit with $s= m/N$ fixed, and use the Gaussian distribution to approximate the Binomial distribution, i.e.,
\bea
    2^{N/2} p_{m,0}^\O(0) \approx \frac2{N} \sqrt{\frac{N}{\pi}} e^{-N(s-1/2)^2}.
\eea
Thus at $t\ll 1/\V$, the second R\'enyi entropy grows linearly,
\bea
	e^{-S_2(t)} = P(t) \approx 1 - \frac{2N\V}{q^2} t \approx e^{-\frac{2N \V}{q^2} t}, \quad S_2(t) = \frac{2N \V}{q^2} t.
\eea
At late time, as we know that the second R\'enyi entropy will saturate at $\frac{N}2 \log 2$, we have thus the following Page curve of second R\'enyi entropy per Majorana fermion,
\bea \label{eq:master_result}
    \S_2(t) \equiv \frac{S_2(t)}{N} = \begin{cases}
                \frac{2\V}{q^2} t, & t < t^* \\
                \frac{1}2 \log 2, & t> t^*
                \end{cases},
\eea
where $t^* = \frac{q^2}{4\V} \log 2$ is the Page time. Even though we start from a different initial density matrix, we can still compare the result from the saddle point solution. We find that two results match exactly for second R\'enyi entropy per Majorana~(\ref{eq:saddle_result}). This is the useful quantity since the Hilbert dimensions are different for the two cases.

We also numerically solve the master equation starting from~(\ref{eq:initial_distribution}) for $q/2=2$. The purity evolution as a function of time is plotted as the solid line in Fig.~\ref{fig:purity}, explicitly showing the Page curve in the Brownian evolution. The dotted line and dash line are given by~(\ref{eq:master_result}). The deviation from the linear increasing dotted line at early time is due to finite $N$ effect. For the dashed line, we have included the finite $N$ correction from $Z_2^f \times Z_2^f$ symmetry. Fig.~\ref{fig:exponential} shows that after the Page time, the approach to the thermal value is given by an exponential function. This is consistent with a $\log N$ scrambling time as discussed in Sec.~\ref{sec:finite_time}, where it is explained by the effect of twist operators.

\section{Conclusion}

We studied the R\'enyi entropy dynamics of coupled Brownian SYK clusters using both path integral and operator dynamics methods. While the Page curve has been observed in random circuit models before, we showed how the replica diagonal and non-diagonal saddle points give rise to the entanglement behavior. This structure is very similar to the replica wormhole scenario obtained in holographic calculations. We also discussed the scrambling time and Page time in Section~\ref{sec:finite_time} and Appendix~\ref{append:twist}, where the crucial effect of the twist operator was discussed. One interesting future direction is to calculate the entanglement entropy directly, and consequently to reveal the role of entanglement islands in more generic quantum mechanical systems.

\section*{Acknowledgements}

We would like to thank Meng Cheng and Pengfei Zhang for useful discussions. S.-K. J. would like to acknowledge helpful discussions with Shenglong Xu in related collaborations. This work is supported by the Simons Foundation via the It From Qubit Collaboration.

\newpage
\appendix

\section{Replica non-diagonal solution} \label{append:non-diagonal}
 
According to the ansatz~(\ref{eq:non-diagonal}) and the Schwinger-Dyson equation~(\ref{eq:SD2}), the self-energy is
\bea
	\hat\Sigma_1 &=&  \tilde f_{1} \left( \ba{cccc} 0 & - \tilde\epsilon^T \\ \tilde \epsilon & 0  \ea \right), \quad \tilde f_1 = \Big(\frac{\J}q f_1(0)^{q-1} + \frac{\V}q f_1(0)^{q/2-1} f_2(0)^{q/2} \Big), \\
	\hat\Sigma_2 &=& \tilde f_2 \left( \ba{cccc} 0 & - 1 \\ 1 & 0  \ea \right), \quad \tilde f_2 = \Big(\frac{\J}q f_2(0)^{q-1} + \frac{\V}q f_2(0)^{q/2-1} f_1(0)^{q/2} \Big),
\eea
where $\tilde f_i$ is a constant. Then~(\ref{eq:SD1}) leads to
\bea
	&&\hat G_1(\omega) = \frac1{\omega^2 + \tilde f_1^2} \left( \ba{cccc} i \omega & - \tilde{f}_1 \tilde\epsilon^T \\ \tilde{f}_1 \tilde \epsilon & - i\omega   \ea \right), \quad \hat G_2(\omega) = \frac1{\omega^2 + \tilde f_2^2} \left( \ba{cccc} i \omega & - \tilde{f}_2 \\ \tilde{f}_2 & - i\omega   \ea \right), \\
	&& \hat G_1 = \frac{e^{-\tilde f_1 |t_{12}|}}2 \left( \ba{cccc} \sgn(t_{12}) & - \tilde\epsilon^T \\ \tilde \epsilon & - \sgn(t_{12})  \ea \right), \hat G_2 = \frac{e^{-\tilde f_2 |t_{12}|}}2 \left( \ba{cccc} \sgn(t_{12}) & -1 \\ 1 & - \sgn(t_{12})  \ea \right).
\eea
Comparing this to the ansatz~(\ref{eq:non-diagonal}), we find the replica non-diagonal solution
\bea
	 f_1(t_{12})= f_2(t_{12}) = e^{- \frac{\J+\V}q |t_{12}|}.
\eea

\section{The solution at finite time} \label{append:finite_T}

We solve the Schwinger-Dyson equation numerically to compare it with the analytic solutions~(\ref{eq:diagonal},~\ref{eq:non-diagonal2}). Starting from a noninteracting solution as an input, we iterate the Schwinger-Dyson equation~(\ref{eq:SD1},~\ref{eq:SD2}) until the result converges. This iteration was used in~\cite{Maldacena:2016remarks} and also in~\cite{Penington:2019replica}.

\begin{figure}
    \centering
\subfigure[]{
    \includegraphics[height=4.5cm]{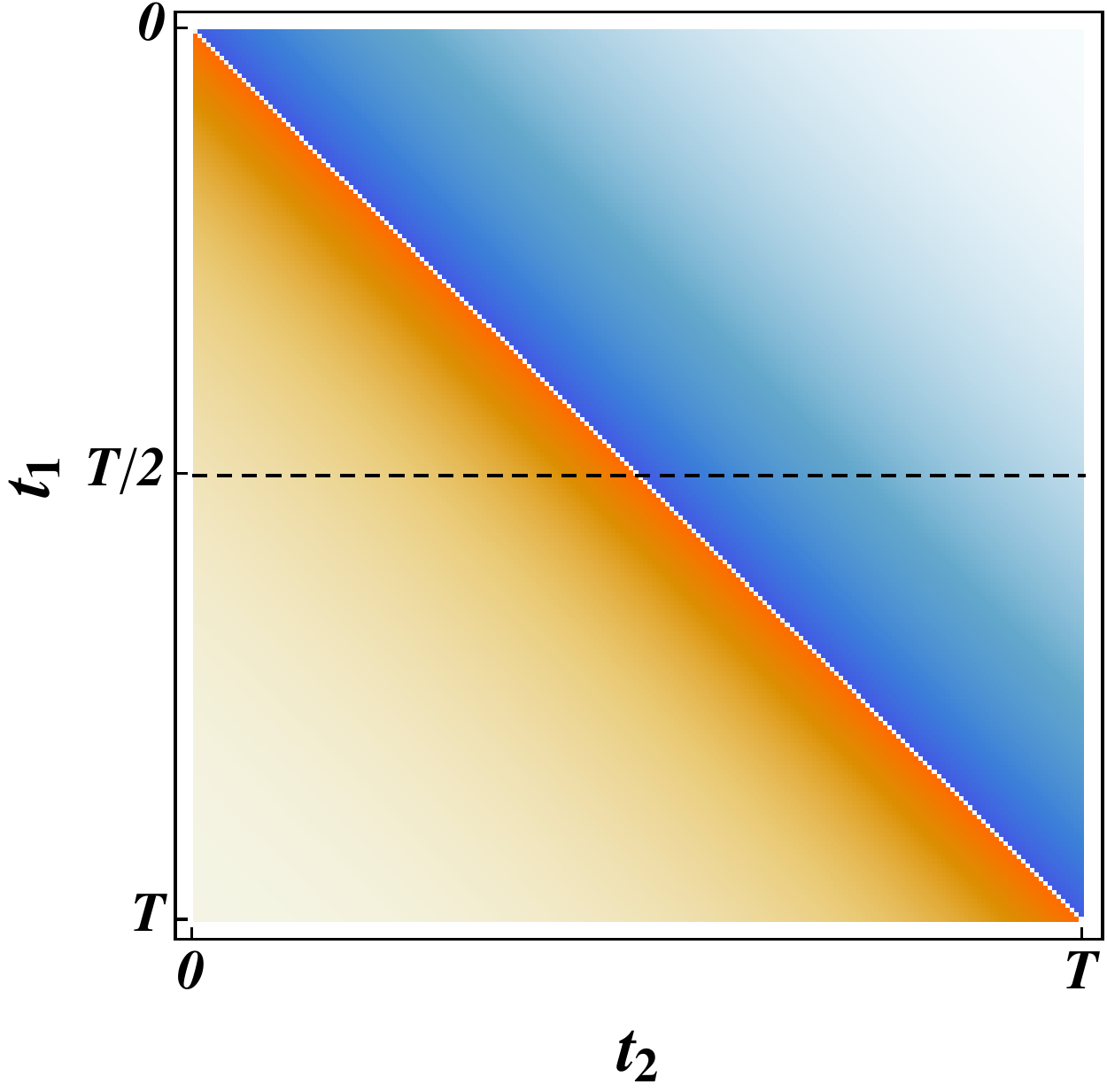}} \qquad \qquad
\subfigure[]{
    \includegraphics[height=4.5cm]{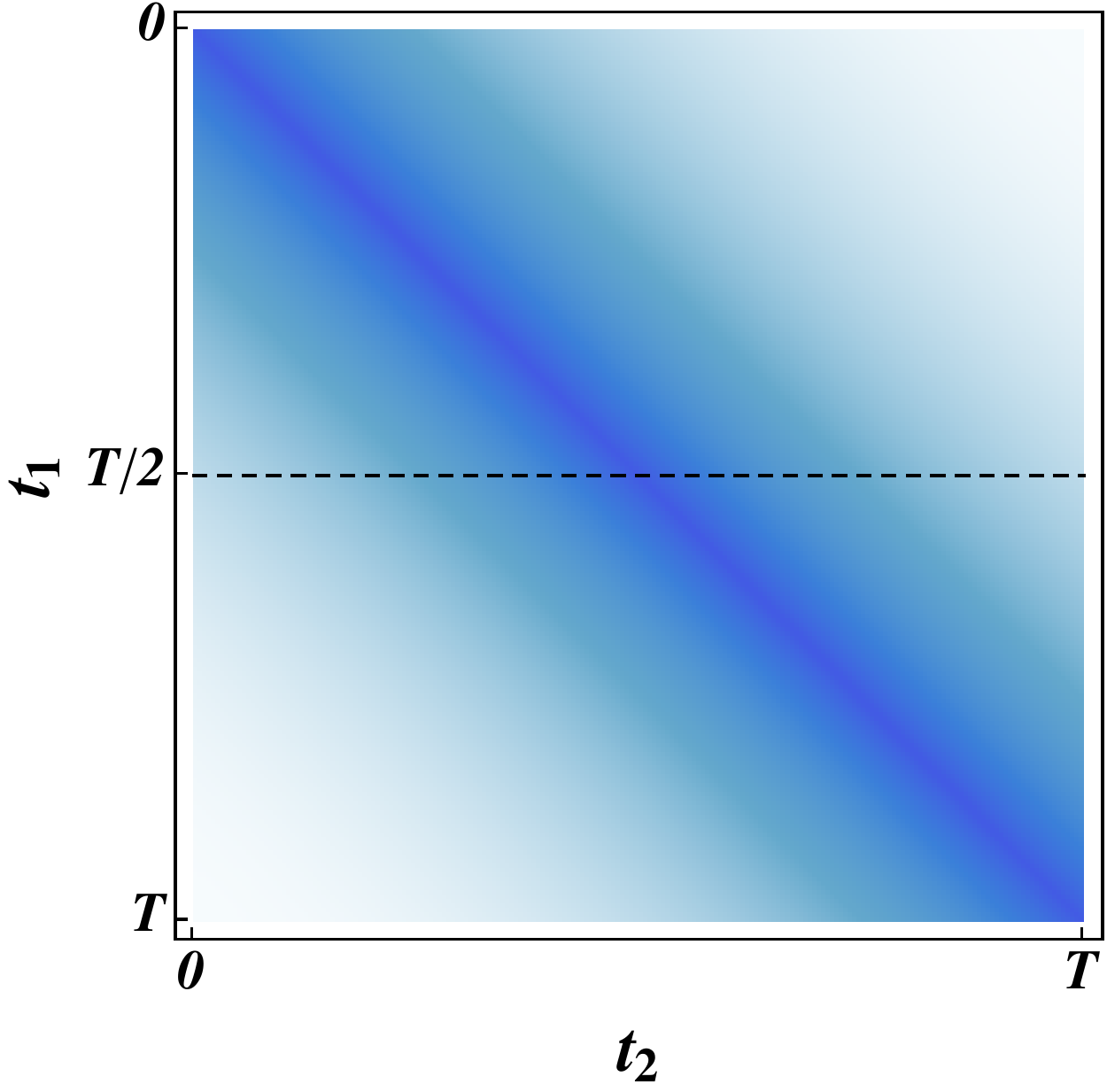}} \\
\subfigure[]{\label{fig:GD_finite_T}
    \includegraphics[height=3cm]{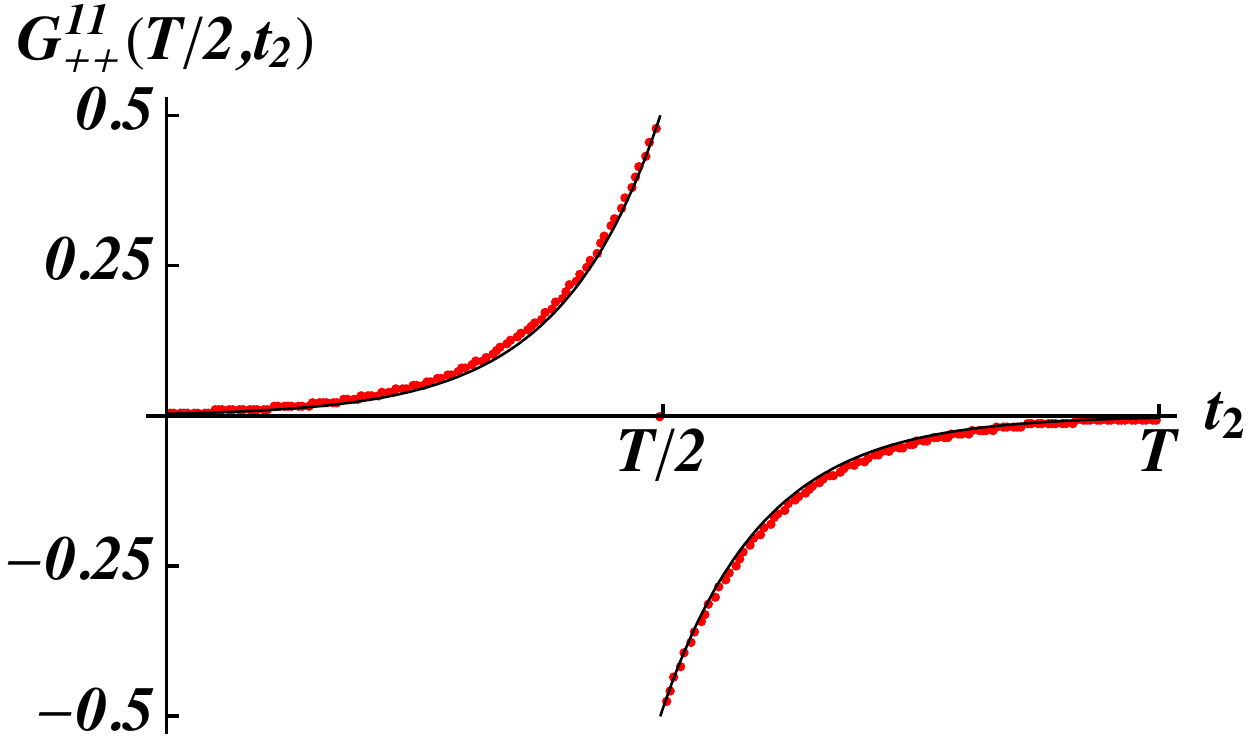}}\qquad
\subfigure[]{\label{fig:GWH_finite_T}
    \includegraphics[height=3cm]{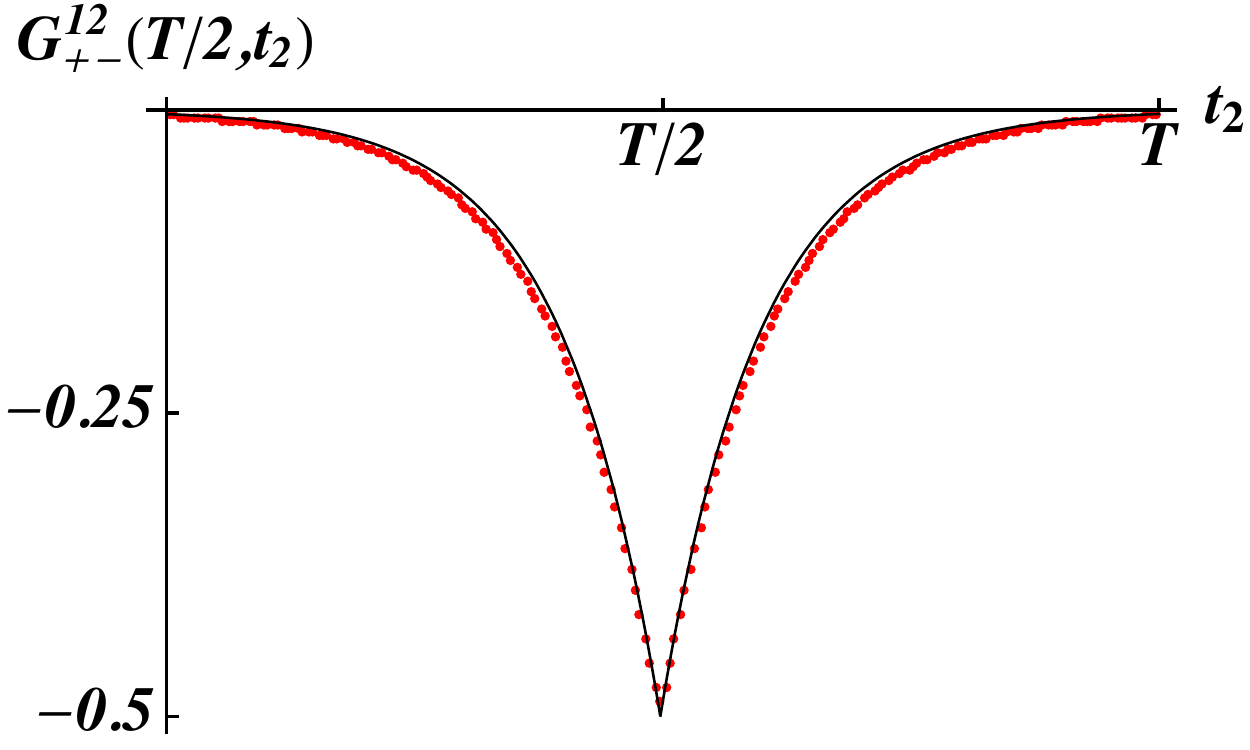}}
    \caption{The replica diagonal (a,c) and replica non-diagonal (b,d) solution of the Schwinger-Dyson equation for $n=2$ R\'enyi entropy. We plot $G_{++}^{11}(t_1, t_2)$ and $G_{+-}^{12}(t_1, t_2)$ for replica diagonal and non-diagonal solutions, respectively. (c,d) are the values of Green's function on the dashed line in (a,b). The red dot (black line) represents the numerical (analytic) solution. We choose the parameter $q=4, \J/\V=1, \J T = 20$. The number of discretization is $M=400$. \label{fig:finite_T}}
\end{figure}

We show the results in Fig.~\ref{fig:finite_T}. The analytic solution~(\ref{eq:diagonal},~\ref{eq:non-diagonal2}) matches the numeric solution quite well for $\J T = 20$ as shown in Fig.~\ref{fig:finite_T} for both diagonal Fig.~\ref{fig:GD_finite_T} and non-diagonal Fig.~\ref{fig:GWH_finite_T} solutions.

\section{Numerical calculation of the saddle point solution and the onshell action} \label{append:onshell}

\begin{figure}
    \centering
    \includegraphics[width=8cm]{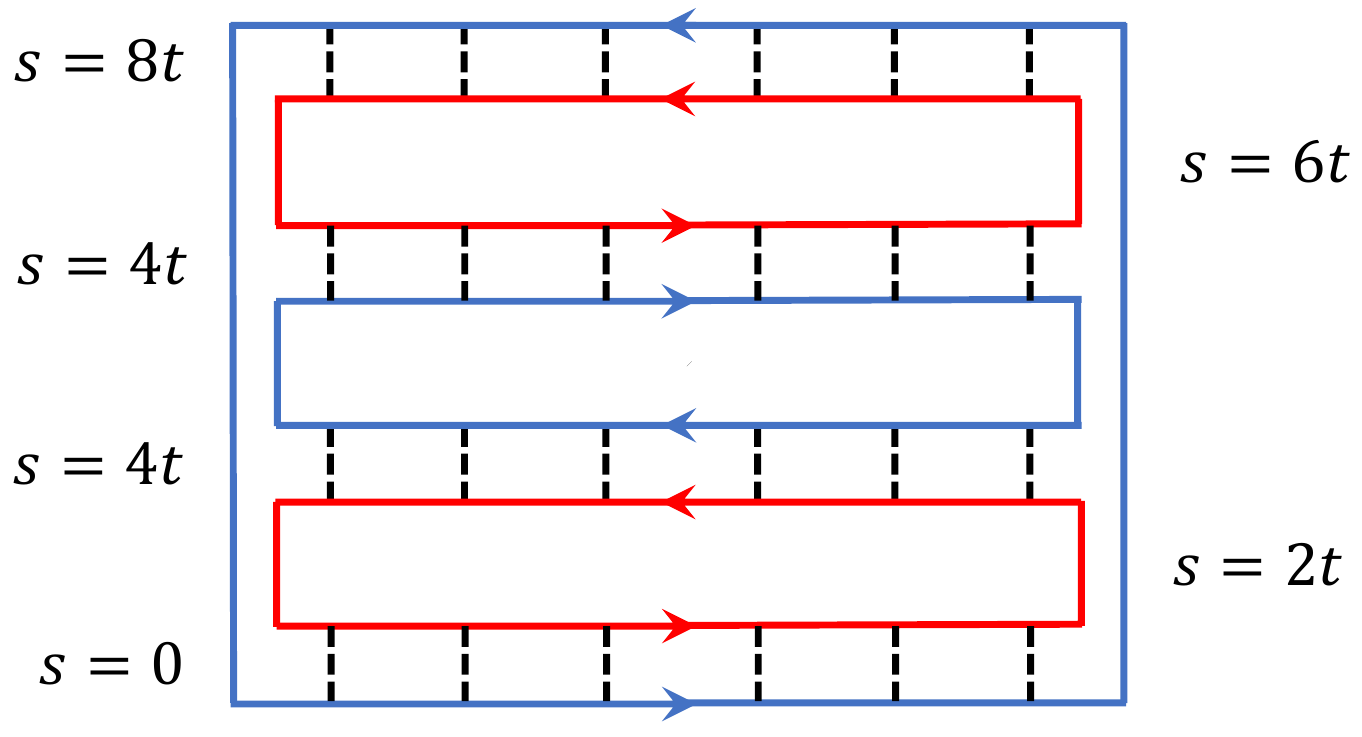}
    \caption{The contour convention of $n=2$ R\'enyi entropy. The blue (red) solid line represents the subsystem $a=1$ ($a=2$). The arrows indicate the direction of the time parametrization. The black dashed lines indicate the interaction between two subsystems.}
    \label{afig:contour}
\end{figure}

For numerical convenience, here we adopt a different convention for the labeling of fields. We put both Keldysh contour indices and replica indices into the time argument $0<s<2 n T$ (a similar convention is used in~\cite{Chen:2020replica}): The forward contour for the $\alpha$-th replica is $s \in ((2\alpha-2)T, (2\alpha-1) T)$, and the backward contour for the $\alpha$-th replica is $s \in ((2\alpha-1)T, 2\alpha T)$. We also introduce a sign factor to capture the forward and backward contour,
\bea
    f(s) = \begin{cases} i, & s \in ((2\alpha-2)T, (2\alpha-1) T) \\
             -i, & s \in ((2\alpha-1)T, 2\alpha T) \end{cases} \qquad
             \alpha = 1,...,n.
\eea
In this convention we also adapt the action such that the interaction between two clusters is local at $s$. As an illustration, the contour convention for $n = 2$ is shown in Fig.~\ref{afig:contour}. In this case, as seen from the figure, a replica diagonal solution in subsystem $a=1$ will have nonvanishing correlation between $s\in (0,2t)$ and $s \in (6t, 8t)$ and between $s\in (2t,4t)$ and $s \in (4t, 6t)$. On the other hand, a nonvanishing correlation between $s\in (0,2t)$ and $s\in(2t,4t)$ and between $s\in (4t,6t)$ and $s \in (6t, 8t)$ for subsystem $a=1$ implies a replica non-diagonal solution. We will assume $a=2$ has a diagonal solution in the following.

In terms of this convention, the action is
\bea
    S &=& \int ds ( \frac12 \psi \partial_s \psi + f(s) H(s) ), \\
    H(s) &=& \sum_{|A|=q, a=1,2} J_{A}^a(s) [\psi_a]_A + \sum_{|A|=|B|=q/2} V_{A,B}(s) [\psi_1]_A [\psi_2]_B.
\eea
where $A=j_1...j_{|A|}$ denotes an ascending list of length $|A|$, and $[\psi_a]_A \equiv i^{|A|/2} \psi_{j_1,a} \psi_{j_2,a} ... \psi_{j_{|A|},a} $ is a short-hand notation for $|A|$-body interaction. The summation is over all possible such lists from $N_a$ Majorana fermions. 

In general, the interaction strength is random variable with possible dependence on time $s$. The distributions of the interactions are defined by vanishing means and the following variances,
\bea
	\overline{J_{A}^a(s)J_{A'}^{a'}(s')} &=& \frac{2^{q-1} q! }{q^2 N_a^{q-1}} \J f_J(s-s') \delta_{A,A'} \delta_{a,a'}, \\
	\overline{V_{A,B}(s)V_{A',B'}(s')} &=& \frac{2^{q} (q/2)!^2}{q^2 N_1^{(q-1)/2} N_2^{(q-1)/2} } \V f_V(s-s') \delta_{A,A'} \delta_{B,B'}, \\
	\delta_{A,A'} &\equiv& \delta_{j_1, j_1'} ... \delta_{j_{|A|}, j_{|A'|}'}.
\eea
where the function $f_J$ and $f_V$ characterize the time dependence of the variances. For Brownian random variable on the contours, $f_J(s,s') = f_V(s,s')= \sum_{\alpha=0}^{n-1} \delta( |s-s'|- 2 \alpha T) + \sum_{\alpha=1}^{2n-1} \delta(s+s'-2\alpha T)] $.
And for regular SYK model, $f_J(s,s') =\J , f_V(s,s') = \V$.

The effective action after averaging over random variables and introducing bilocal fields, i.e. the Green's function $G_a(s_1,s_2) = \frac1{N_a} \sum_{j=1}^{N_a} \psi_{j,a}(s) \psi_{j,a}(s')$ and the self-energy $\Sigma_a(s,s')$, reads
\bea
    -I &=& \sum_a N_a \Big[  \frac12 \Tr \log(G_{0,a}^{-1} - \Sigma_a) - \frac12 \int ds_1 ds_2 \Sigma_a(s_1,s_2) G_a(s_1,s_2) \\
    && + \frac{\J}{4q^2} \int ds_1 ds_2 f_J(s_1,s_2) f(s_1) f(s_2) (2G_a(s_1,s_2))^q \Big] \nn\\
    && + \sqrt{N_1 N_2} \frac{\V}{2q^2} \int ds_1 ds_2 f_V(s_1,s_2) f(s_1) f(s_2) (2G_1(s_1,s_2))^{q/2} (2G_2(s_1,s_2))^{q/2},
\eea
The Schwinger-Dyson equation follows from the effective action is given by
\bea
\label{aeq:SD1}    \hat G^{-1}_a &=& \hat G_{0,a}^{-1} - \hat \Sigma_a, \\
\label{aeq:SD2}    \Sigma_a(s_1, s_2) &=& \frac{\J}{q} f_J(s_1,s_2) f(s_1) f(s_2) (2G_a(s_1,s_2))^{q-1} \nn\\
    && + \sqrt{\frac{N_{\bar a}}{N_a}}\frac{\V}{q} f_V(s_1,s_2) f(s_1) f(s_2) (2G_a(s_1,s_2))^{q/2-1} (2G_{\bar a}(s_1, s_2))^{q/2}.
\eea

\begin{figure}
    \centering
\subfigure[]{
    \includegraphics[width=5cm]{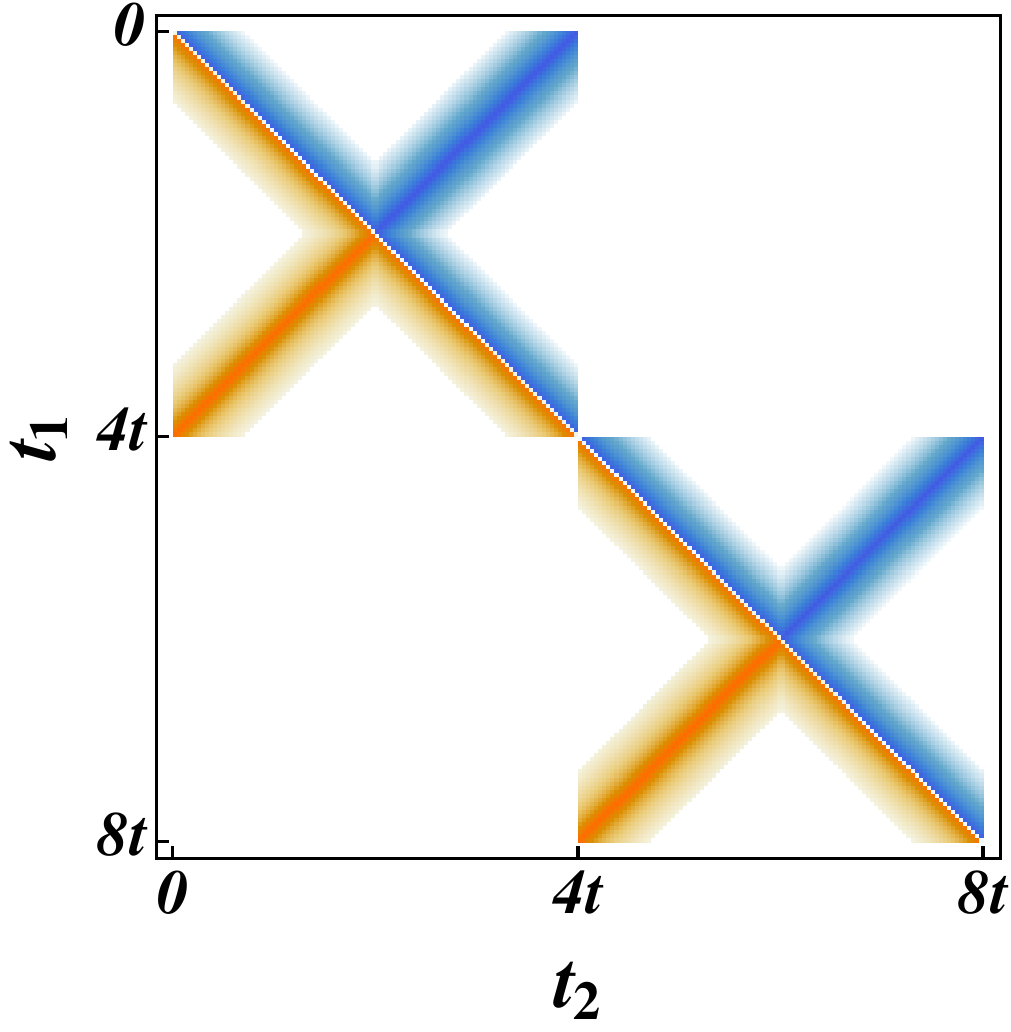}} \qquad
\subfigure[]{\label{fig:GWH}
    \includegraphics[width=5cm]{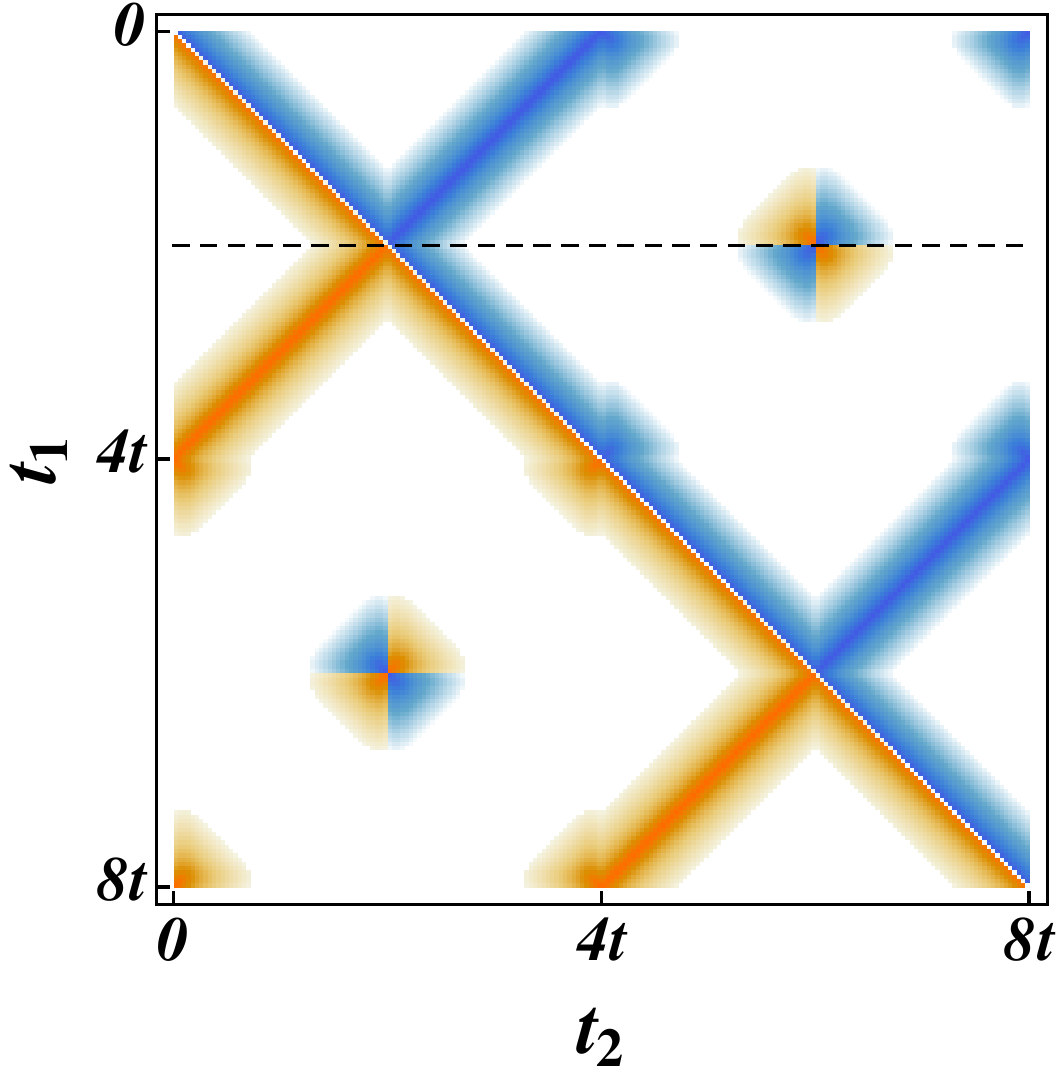}} \\
\subfigure[]{
    \includegraphics[width=5cm]{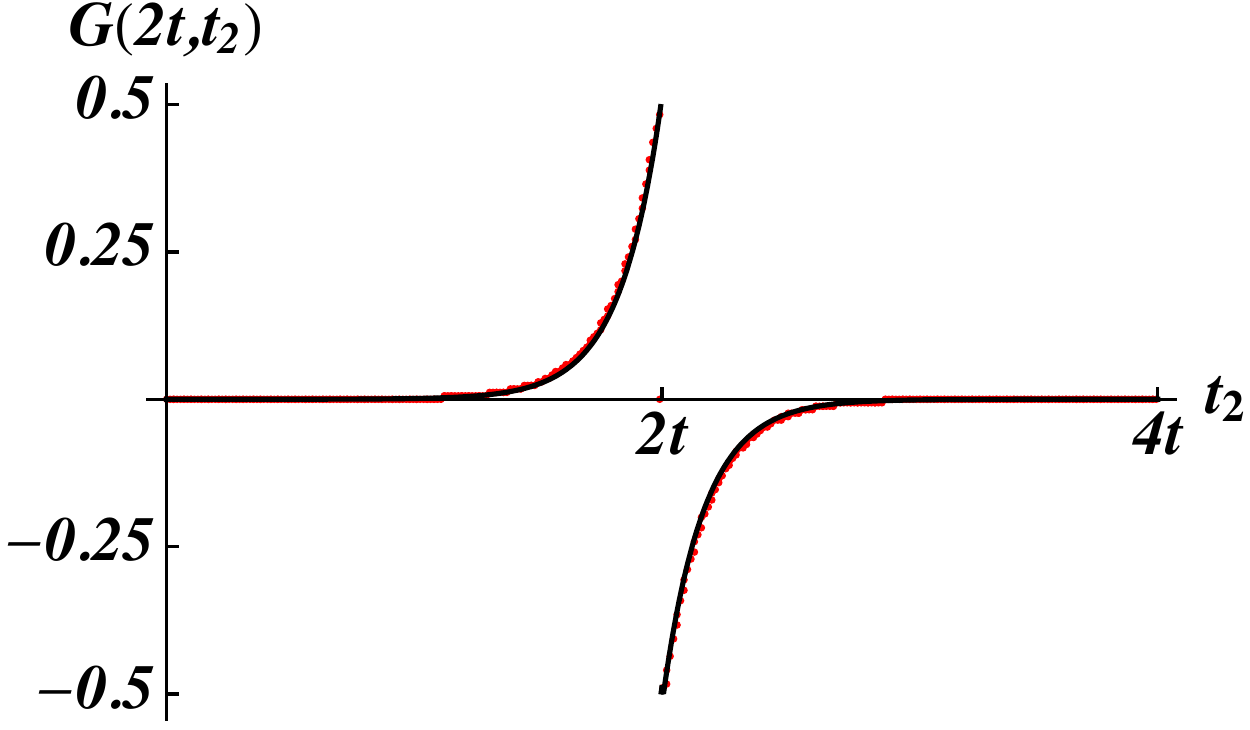}} \qquad
\subfigure[]{
    \includegraphics[width=5cm]{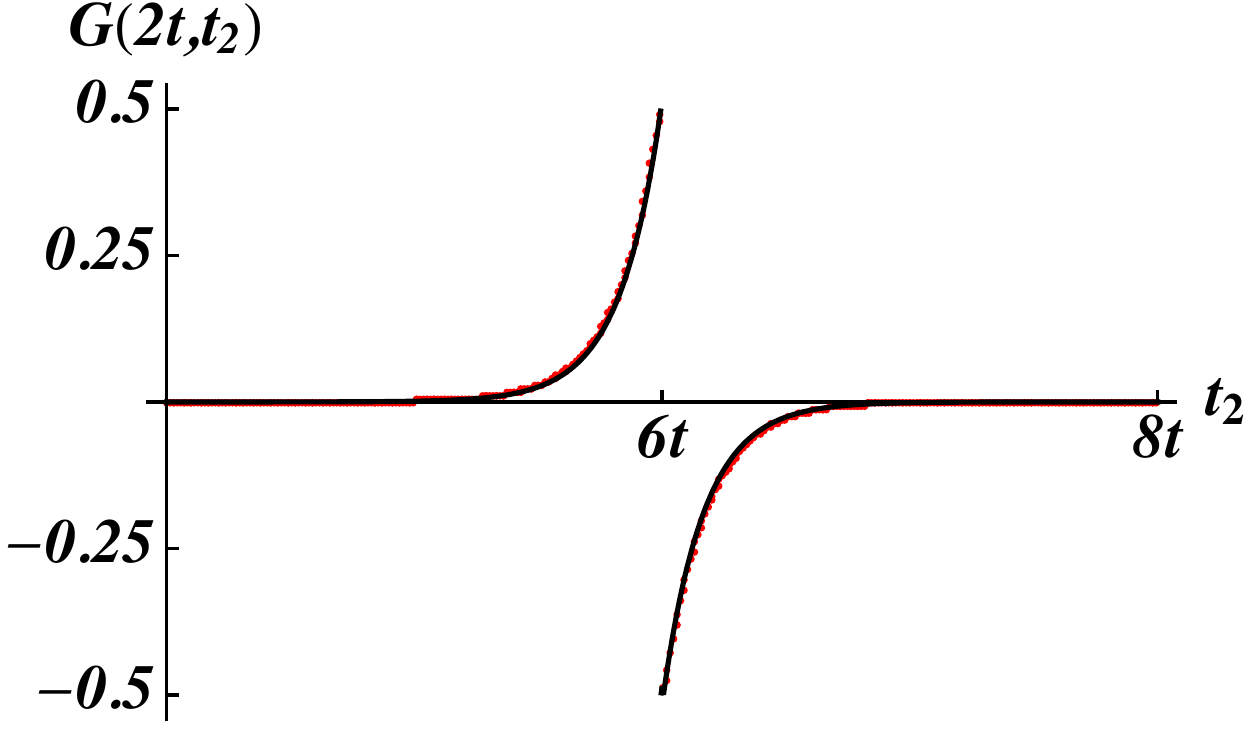}}
    \caption{The replica diagonal solution $G_2$ (a) and replica non-diagonal solution $G_1$ (b) of Schwinger-Dyson equation for the second R\'enyi entropy. (c,d) Comparison between numerical solutions and analytic solutions located on the dashed line in (b). (d) The nonvanishing correlation sourced by the twist operator is also given by the profile of $e^{-(\J+\V)|t|/q }$ at the dashed line in (b) due to the boundary condition at $t_1 = 2t$. We choose the parameter $\J T =20$, $\V/\J =0.2$, $q=4$. Note that $T=2t$. The number of discretization is $M=400$.  \label{fig:G2_numerical}}
\end{figure}
\begin{figure}
    \centering
\subfigure[]{
    \includegraphics[width=5cm]{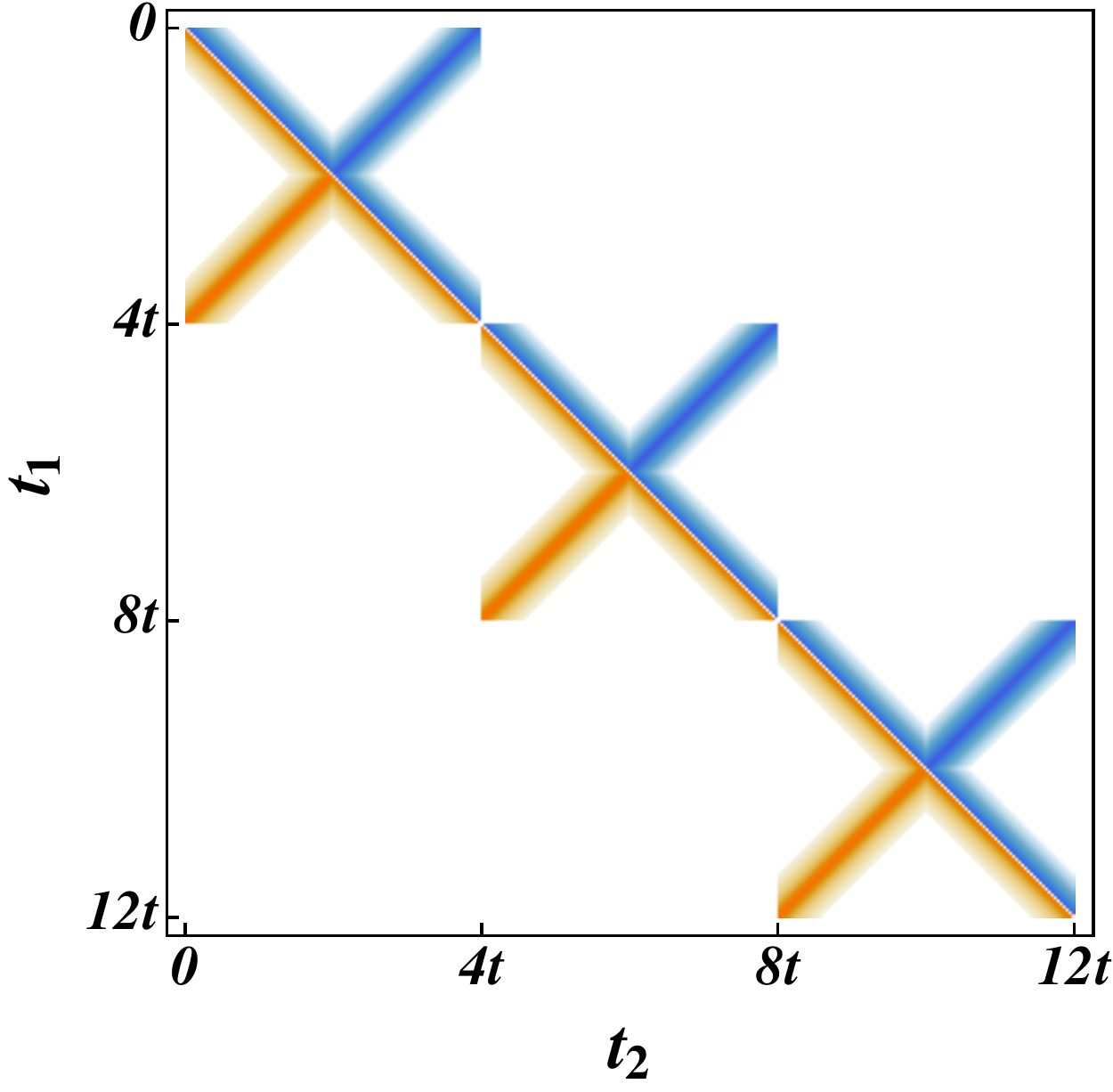}} \qquad
\subfigure[]{\label{fig:GWH3_Append}
    \includegraphics[width=5cm]{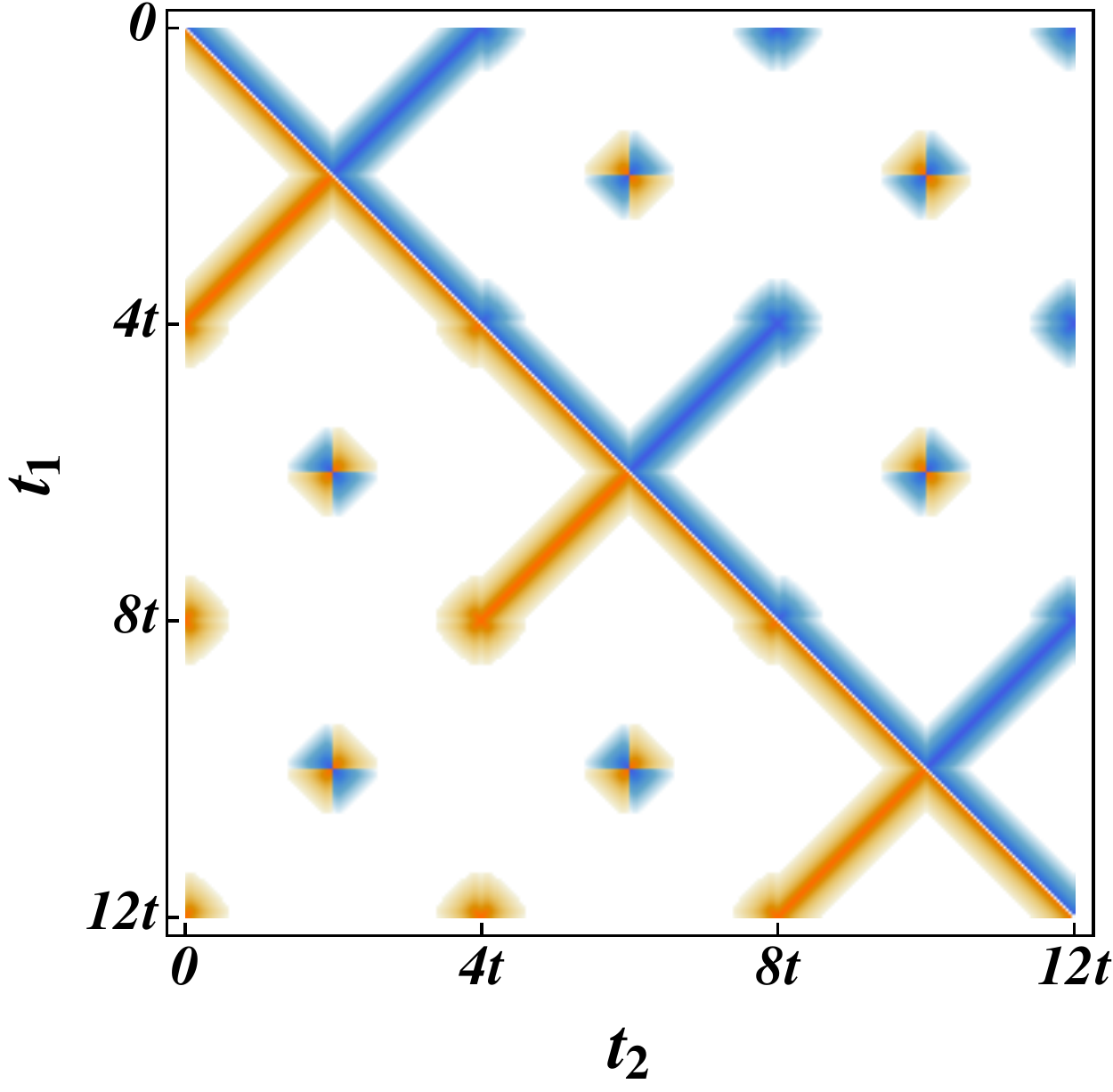}}
    \caption{The replica diagonal solution $G_2$ (a) and replica non-diagonal solution $G_1$ (b) of Schwinger-Dyson equation for the third R\'enyi entropy. (b) shows clearly the nonvanishing correlation is sourced by the twist operator. We choose the parameter $\J T = 24$, $\V/\J=0.2$, $q=4$. The number of discretization is $M=400$.   \label{fig:G3_numerical}}
\end{figure}

For simplicity, we consider $N_1 = N_2 = N$. We numerically solve the Schwinger-Dyson equation (\ref{aeq:SD1},~\ref{aeq:SD2}) for $n=2,3$ and look for replica non-diagonal solution. In doing so, we use $ G_{0,a}(s,s') = \frac12 \sgn(s-s')$ for times located at the same close time path. An illustration of the close time bath for $n=2$ is given in Fig.~\ref{afig:contour}. To get the replica non-diagonal solution, we start from an initial ansatz with small but non-zero non-diagonal correlations for subsystem $a=1$ and a diagonal initial ansatz for subsystem $a=2$. The results for $n=2$ are shown in Fig.~\ref{fig:G2_numerical}. As we discuss in above, Fig.~\ref{fig:GWH} is a replica non-diagonal solution. It is intuitive to note from the figures that the only difference between the replica diagonal solution and the replica non-diagonal solution is those nonvanishing correlations at $\{ 4t<s_1<8t \} \cap \{ 0<s_2<4t \}$ and $\{ 0<t_1<4t \} \cap \{ 4t<t_2<8t \}$ sourced by the twist operators located at $s=0,2t,6t,8t$. We also get the results for $n=3$ R\'enyi entropy, which are shown in Fig.~\ref{fig:GWH3_Append}, and there are six twist operators.

\begin{figure}
    \centering
    \includegraphics[width=6cm]{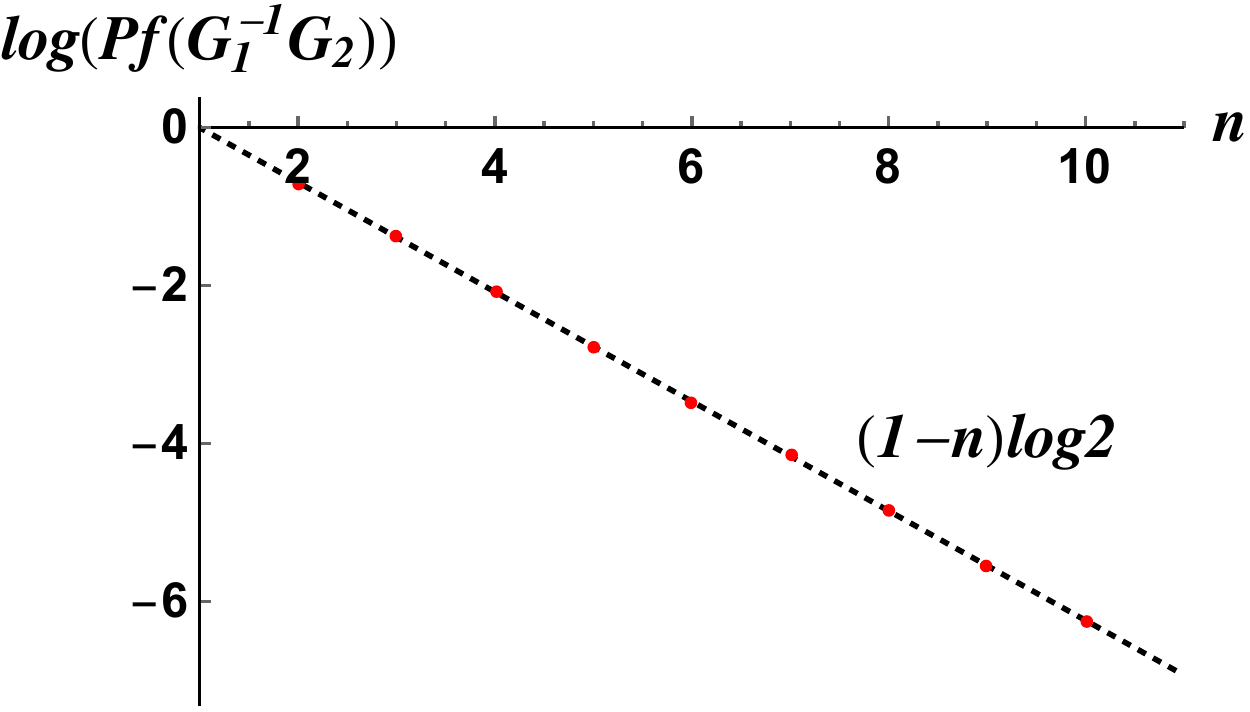}
    \caption{The Pfaffian in onshell action as a function of $n$. We choose $\J T= 50$, $\V/\J = 0.2$, $q=4$. The number of discretization is $M=40$. \label{fig:twist}}
\end{figure}

We also numerically calculate the Pfaffian in the calculation of onshell action and check the validity of~(\ref{eq:page}). The result is plotted in Fig.~\ref{fig:twist} where we calculate $\log \Pf( G_1^{-1} G_2)$ for the R\'enyi entropy $n=2,...,10$. The dashed line is $(1-n) \log 2$, and we find excellent agreement of the numerical evaluated values and~(\ref{eq:page}).

\section{Non-diagonal solutions and twist operators}\label{append:twist}

In this section, we discuss the effect of the twist operators. We will focus on the replica non-diagonal solution for the second R\'eny entropy $n=2$ for simplicity, while the generalization to other R\'enyi entropy is straightforward. The first equation (\ref{aeq:SD1}) couples functions non-locally in time domain, while the second equation (\ref{aeq:SD2}) is local. Using the large-$q$ ansatz, $G_a = g_{a0} (1+ \frac{g_a}q)$, the Schwinger-Dyson equation becomes,
\bea
    \Sigma_a(s_1,s_2) &=&  - \frac1q \partial_{s_1} \partial_{s_2} g_{a0} g_a(s_1, s_2), \\
    \Sigma_a(s_1, s_2) &=& \frac{\J}{q} f_J(s_1,s_2) f(s_1) f(s_2) (2g_{a0})^{q-1} e^{g_a(s_1,s_2)} \nn\\
    && + \sqrt{\frac{N_{\bar a}}{N_a}} \frac{\V}q f_V(s_1,s_2) f(s_1) f(s_2) (2g_{a0})^{q/2-1} (2g_{\bar a0})^{q/2} e^{\frac12 (g_a(s_1,s_2) + g_{\bar a}(s_1, s_2))}.
\eea
One advantage of the large-$q$ equation of motion is that it becomes local in time variables. For simplicity, we will assume $N_1=N_2=N$. We also assume $g_{20} = \frac12 \sgn(s_1-s_2)$ when $s_1, s_2$ are located at the same close time path and zero otherwise, while $g_{10} = \frac12 \sgn(s_1-s_2)$ to look for the non-diagonal solutions.

We can solve it in the regime $0<s_1<T$ and $0<s_2<T$. For Brownian case, the large-$q$ saddle point equation reads,
\bea
    \partial_{s_1} \partial_{s_2} g_a(s_1, s_2) &=& 2 \J \delta(s_1 - s_2) e^{g_a(s_1, s_2)} + 2 \V \delta(s_1-s_2) e^{\frac12 (g_a(s_1, s_2) + g_{\bar a}(s_1, s_2))}.
\eea
The equation can be solved by realizing it is $\partial_{s_1} \partial_{s_2} g_a(s_1, s_2) = 0$ when $s_1 \ne s_2$, and the $\delta$ function leads to a jump in the first derivative at $ s = s_1 = s_2$ that can be solved easily. Supplementing with the boundary condition $g_a(s,s) = 0$ and $g_a(s_1,s_2) = g_a(s_2,s_1)$, the solutions are
\bea
    g_1(s_1, s_2) = g_2 (s_1, s_2) = - (\J + \V) |s_1-s_2|, \quad \{ 0<s_1<T, 0<s_2<T \}.
\eea
We can extend such calculations to the regime $0<s_1<2T$ and $0<s_2<2T$, 
\bea\label{eq:diagonal-G2}
    G_a(s_1,s_2) = \frac{\sgn(s_1-s_2)}2 \begin{cases} e^{- \frac{\J + \V}q |s_1-s_2|} , \quad & \{s_1,s_2\} \in \{(0,T),(0,T) \} \cup \{(T,2T),(T,2T) \}  \\
    e^{- \frac{\J + \V}q |T-s_1-s_2|}, \quad & \text{otherwise} \end{cases}.
\eea
This is consistent with the analytic solution~(\ref{eq:non-diagonal2}) and the numeric solution shown in Fig.~(\ref{fig:G2_numerical}). More generally, the solutions in $(\alpha-1)2T<s_1,s_2< 2\alpha T$, $\alpha = 1,...,n$ will be the same.

To investigate the nonvanishing correlation induced by the twist operator near the boundary, we first focus on the regime $3T<s_1<4T$ and $0<s_2<T$. In this regime, there are two twist operators, where $T_1$ locates at $(4T,0)$ and $T_2$ locates $(3T,T)$ as also indicated by the nonvanishing correlations in Fig.~\ref{fig:G2_numerical}. So the boundary conditions are $\psi(T)= \psi(3T)$ and $\psi(0)= - \psi(4T)$. The minus sign is because of the Fermi operator. To simplify the notation, we shift $s_1 \rightarrow s_1+3T$, so the regime is $0<s_1,s_2<T$. After the redefinition, the large-$q$ equation of motion in this regime is
\bea\label{aeq:large-q}
    \partial_{s_1} \partial_{s_2} g_1(s_1,s_2) = -2 \J \delta(T-s_1-s_2) e^{g_1(s_1, s_2)}.
\eea
The absence of the $\V$ term is because the replica diagonal solution of subsystem $a=2$ vanishes in this regime.

The solution is exponentially suppressed at large $s$, and at the large time, i.e., $\J T \gg 1$, the two twist operators are separated by a large distance $T$. So we can further assume in this regime the induced solution is separable as follows,
\bea
    G_1(s_1, s_2) = G_{T_1}(s_1, s_2) + G_{T_2}(s_1, s_2),
\eea
where $G_{T_i}$ denotes the induced solution by twist operator $T_i$. At large-$q$ limit, $G_{T_i} \approx (1+ \frac{g_{T_i}}{q} ) $, and $g_{T_i}$ satisfies large-$q$ equation of motion~(\ref{aeq:large-q}). But they satisfy different boundary conditions because the two twist operators locate at different places, i.e.,
\bea
    g_{T_1}(s_1,0) = - (\J + \V) (T-s_1), \quad g_{T_1}(T,s_2)= -(\J + \V) s_2, \\
    g_{T_2}(s_1,T) = - (\J + \V) s_1, \quad g_{T_2}(0,s_2)= -(\J + \V)(T-s_2).
\eea
Let us first look at $g_{T_1}$. Owing to the delta function in the right-hand-side of~(\ref{aeq:large-q}), the solution is not differentiable at $s_1+s_2 = T$, so we assume
\bea
g_{T_1}(s_1,s_2) = \begin{cases} g_I(s_1,s_2), & s_1 + s_2 \le T \\
                            g_{II}(s_1, s_2), & s_1 + s_2 \ge T
                \end{cases}.
\eea
Taking the boundary conditions into consideration, the solution has the form,
\bea
    g_I(s_1, s_2) = - (\J + \V)(T-s_1) + f_{T_1}(T-s_2), \quad g_{II}(s_1, s_2) = - (\J + \V)s_2 + f_{T_1}(s_1),
\eea
and $f_{T_1}$ satisfies
\bea
    \partial_s f_{T_1}(s) = \J + \V -2 \J e^{- (\V+ \J)(T-s)+ f_{T_1}(s)}, \quad f_{T_1}(T) = 0.
\eea

It is not hard to solve above differential equation, which leads to the solution,
\bea
    f_{T_1}(s) = -(\J + \V)(T-s) - \log \frac{e^{-2(\J + \V)(T-s)}\J + \V}{\J + \V}.
\eea
And consequently, one can get the correlation function $G_{T_1} \approx \frac12 e^{g_{T_1}/q}$ induced by the twist operator $T_1$, 
\bea
    g_{T_1}(s_1,s_2) = \begin{cases} -(\J + \V)(T-s_1-s_2)- \log  \frac{e^{-2(\J + \V)s_2}\J + \V}{\J + \V}, & s_1 + s_2 \le T \\
                            -(\J + \V)(T-s_1-s_2)- \log  \frac{e^{-2(\J + \V)(T-s_1)}\J + \V}{\J + \V}, & s_1 + s_2 \ge T
                \end{cases}.
\eea

To simplify the notation, we define the induced Green's function as
\bea\label{aeq:twist}
    G_{T}(s_1, s_2) = \frac12 \exp \Big[- \frac1q \Big( (\J + \V)(s_1+s_2)+ \log  \frac{e^{-2(\J + \V)\min(s_1,s_2)}\J + \V}{\J + \V} \Big) \Big].
\eea
Intuitively, this function is exponentially suppressed away from $s_1 = s_2 = 0$ where the twist operator supposed to be located. So the solution at the regime is
\bea
    G_1(s_1,s_2) = G_T(T-s_1, s_2) + G_T(s_1, T-s_2).
\eea
This is an approximate solution with accuracy $O(e^{ -\J T})$. The solutions in other regimes can be obtained in the same way, so we do not have to detail the calculation. 

Now we can discuss the effect of these twist operators to the onshell action. We mainly discuss the second R\'enyi entropy, but the results can be extended to $n$-th R\'enyi entropy by small modifications. Taking into account the twist operators, the replica non-diagonal solution is $G_1 = G_2 + G_{T_1} + G_{T_2}$, where $G_2$ denotes the diagonal solution in subsystem $a=2$~(\ref{eq:diagonal-G2}), and $G_{T_1}$ and $G_{T_2}$ are induced solutions by the twist operators $T_1$ and $T_2$, respectively. Notice $G_2$ and $G_{T_i}$ have different domain of support. The onshell action is
\bea
    \log \frac{e^{-I^{(2)}}}{Z^2} &=& \frac{N}2 \Tr(\log(G_1^{-1} G_2)) + N \frac{1-q}{q^2} \J \int f(s_1) f(s_2) f_\J(s_1, s_2)(G_{T_1} + G_{T_2})^q \\
    &\approx& - N \log 2 -\frac{N}2 \int d s_1 d s_2 \partial_{s_1} G_{T_1}(s_1, s_2) \partial_{s_2} G_{T_2}(s_2, s_1)
\eea
where in the second line, we expand the $\Tr \log$ term and keep the lowest-order coupling between two twist operators $G_{T_1}$ and $ G_{T_2}$, because the factorized part contributes to the coarse grained entropy $- N \log 2$~\cite{Chen:2020replica}, and we neglect other subleading terms in the large-$q$ limit. So including the parts from coupled induced Green's function, the contribution from the replica non-diagonal solution is
\bea
    \log \frac{e^{-I^{(2)}}}{Z^2} \approx - N \log 2 + \frac{N}{q^2} (\J + \V)^2 T^2 e^{- \frac{2(\J + \V)T}q}.
\eea
In getting above results, we neglect the second term in~(\ref{aeq:twist}) which will not change the essential exponential factor. Then the second R\'enyi entropy from two saddle points reads
\bea
    e^{-S_2(T)} &=&  \frac{e^{-I^{(1)}}+ e^{-I^{(2)}}}{Z^2} = e^{-\frac{ 2 N \V T}{q^2}} + e^{- N \log 2 + \frac{N}{q^2} (\J + \V)^2 T^2 e^{- \frac{2(\J + \V)T}q}}.
\eea
After the Page time when the replica non-diagonal saddle point dominates, the R\'enyi entropy is actually not independent of time. The exponentially small overlaps between two twist operators mean that it takes times proportional to $\log N$ to fully scramble the information~\cite{Lashkari:2011towards, Gharibyan:2018onset}.

The large-$q$ analysis of the twist operator can be extended to the regular SYK model. Here we calculate it at the infinite temperature for an illustration. The solutions at diagonal part is simple, yielding the solution
\bea
    G_a(s_1,s_2)= \begin{cases} \frac{\sgn(s_1-s_2)}2\Big(\frac1{\cosh(\J_0 |s_1-s_2|)}\Big)^{2/q} ,  \quad & \{s_1,s_2\} \in \{(0,T),(0,T) \} \cup \{(T,2T),(T,2T) \}\\
                                \frac{\sgn(s_1-s_2)}2 \Big(\frac1{\cosh(\J_0 |T-s_1-s_2|)}\Big)^{2/q}, \quad & \text{otherwise}
                \end{cases}.
\eea
where $\J_0 = \sqrt{\J^2 + \V^2}$. So let us focus again on the regime $3T<s_1<4T$ and $0<s_2<T$. Redefining $s_1 \rightarrow s_1+3T$, the regime is $0<s_1,s_2<T$. The equation of motion now reads
\bea
    \partial_{s_1} \partial_{s_2} g_1(s_1,s_2) = -2 \J^2 e^{g_1(s_1, s_2)}.
\eea
The absence of the $\V$ term is because the replica diagonal solution of subsystem $a=2$ vanishes in this regime. A general solution to above Liouville equation is $g_1(s_1,s_2) = \log \frac{h_1'(s_1) h_2'(s_2)}{\J^2 (h_1(s_1)- h_2(s_2)) }$. We expect the solution is exponentially suppressed at large $s$, and at the large time, i.e., $\J T \gg 1$, the two twist operators are separated by a large distance $T$. So we can further assume in this regime the induced solution is separable as follows,
\bea
    G_1(s_1, s_2) = G_{T_1}(s_1, s_2) + G_{T_2}(s_1, s_2),
\eea
where $G_{T_i}$ denotes the induced solution by twist operator $T_i$. At large-$q$ limit, $G_{T_i} \approx (1+ \frac{g_{T_i}}{q} ) $, and $g_{T_i}$ satisfies the Liouville equation. But they satisfy different boundary conditions because the two twist operators locate at different places, i.e.,
\bea
    g_{T_1}(s_1,0) = 2\log \frac1{\cosh \J_0(T-s_1)}, \quad g_{T_1}(T,s_2)= 2\log\frac1{\cosh \J_0 s_2}, \\
    g_{T_2}(s_1,T) =  2\log \frac1{\cosh \J_0 s_1}, \quad g_{T_2}(0,s_2)=  2\log \frac1{\cosh \J_0(T-s_2)}.
\eea
After we take into account the boundary conditions, it is straightforward to get the following solutions,
\bea
   && G_{T_1}(s_1,s_2) = G_T(T-s_1, s_2), \quad G_{T_2} = G_T(s_1, T-s_2), \\
    && G_T(s_1, s_2) = \frac12 \frac{1}{(\cosh \J_0 s_1  \cosh \J_0 s_2 + \frac{\J^2}{\J_0^2} \sinh \J_0 s_1  \sinh \J_0 s_2 )^2}.
\eea
It will interesting to explore the effect of these twist operators in more details, which we leave as a future work.

\section{Derivation of the master equation} \label{append:master}

We derive the master equation in this section. We start from~(\ref{eq:pmmt}). Using the properties of the Majorana basis, when $\Gamma_{C,D}^{(q)} $ and $\Gamma_{A,0}$ share even (odd) Majorana operators, it leads to a positive (negative) sign in the following,
\bea
	\Gamma_{C,D}^{(q)} \Gamma_{A,B} \Gamma_{C,D}^{(q)} = \Gamma_{A,B}, \quad \text{if the sum of number of common elements in $A$, $C$ and $B$, $D$ is even} \nn \\
	\Gamma_{C,D}^{(q)} \Gamma_{A,B} \Gamma_{C,D}^{(q)} = - \Gamma_{A,B}, \quad \text{if the sum of number of common elements in $A$, $C$ and $B$, $D$ is odd} \nn
\eea

For a fixed list $A$ ($B$), if $C$ ($D$) and $A$ ($B$) have $k$ ($k'$) common elements, the number in the summation over $C$ is given by $C_{N-m}^{q-k} C_m^k$ ($C_{N-m'}^{q-k'} C_{m'}^{k'}$) and $C_{N-m}^{q/2-k} C_{m}^k$ ($C_{N-m'}^{q/2-k'} C_{m'}^{k'}$) for the intra and inter subsystem interactions, respectively. So the second line in~(\ref{eq:pmm1}) becomes
\bea
	&& 2^{-2N} \sum_{|A|=m, |B|=m'} 2 \Tr[\O(t) \Gamma_{A,B}] \sum_{C,D}\sigma_{C,D} \Tr[\O(t) \Gamma_{C,D}^{(q)} \Gamma_{A,B} \Gamma_{C,D}^{(q)}] \\ 
	&=& 2 \Big[\sum_{k=0}^{\min(q,m)} (-1)^k C_{N-m}^{q-k} C_{m}^k  \sigma_{0}+\sum_{k'=0}^{\min(q,m')} (-1)^{k'} C_{N-m'}^{q-k'} C_{m'}^{k'} \sigma_{0} \nn\\
	&& +  \sum_{k=0}^{\min{(q/2,m)}} (-1)^{k} C_{N-m}^{q/2-k} C_{m}^k \sum_{k'=0}^{\min{(q/2,m')}} (-1)^{k'}  C_{N-m'}^{q/2-k'} C_{m'}^{k'} \sigma_{1} \Big] p_{m,m'}(t).
\eea

One can combine it with the first line in~(\ref{eq:pmm1}) to give the outgoing rate,
\bea
	&& - 2( 2 C_N^q \sigma_{0} + (C_N^{q/2})^2 \sigma_{1} ) p_{m,m'}(t) + 2^{-2N} \sum_{A,B,C,D} 2  \Tr[\O(t) \Gamma_{A,B}]\sigma_{C,D} \Tr[\O(t) \Gamma_{C,D}^q \Gamma_{A,B} \Gamma_{C,D}^q] \nn \\
	&=& -4 \Big[ \sigma_{0} \sum_{k=1, \text{odd}}^{\min(q,m)} C_{N-m}^{q-k} C_{m}^k + \sigma_{0} \sum_{k'=1, \text{odd}}^{\min(q,m')} C_{N-m'}^{q-k'} C_{m'}^{k'}  \nn\\
	&& +  \sigma_{1} \sum_{k=0}^{\min(q/2,m)} \sum_{k'=0}^{\min(q/2,m')} \frac{1-(-1)^{k+k'}}2C_{N-m}^{q/2-k} C_{m}^k  C_{N-m'}^{q/2-k'} C_{m'}^{k'} \Big] p_{m,m'}(t).
\eea
In deriving the result we have used the combinatorial identity $\sum_{k=0}^{\min(m, q)} C_{m}^k C_n^{q-k} = C_{m+n}^q$.

We now calculate the third line in~(\ref{eq:pmm1}). The commutator $[\Gamma_{A,B} ,\Gamma_{C,D}^{(q)}]$ vanishes unless $\Gamma_{A,B}$ and $\Gamma_{C,D}^{(q)}$ shares odd common Majorana operators. Assuming lists $A$ and $C$ have $k$  common elements, the length of $[\Gamma_{A,B} ,\Gamma_{C,D}^{(q)}]$ is $m + |C| - 2k + |D|$. Here, $|C|=q, q/2$ for intra and inter subsystem interactions, respectively. The summation over $|A|=m, |C|$ overcounts the number of terms in $\{ E, |E|=m+|C|-2k \}$. The overcounting factor comes from the number of ways to decompose $m+|C|-2k$ length list into two lists $A$ and $C$ with $k$ common elements, i.e.,	$C_{N-(m+|C|-2k)}^k C_{m+|C|-2k}^{m-k}$. One can make similar analysis to list $B$ and $D$. Then the third line in~(\ref{eq:pmm1}) gives rise to the incoming rate,
\bea
	&&  - 2^{-2N} \sum_{|A|=m,|B|=m'}  \sum_{C,D} \sigma_{C,D}  \Tr^2 (\O(t) [\Gamma_{A,B} ,\Gamma_{C,D}^{(q)} ])\\
	&=& 4 \Big[ \sigma_0 \sum_{k=1, \text{odd}}^{\min(q,m)}  C_{m+q-2k}^{m-k} C_{N-(m+q-2k)}^k p_{m+q-2k,m'}(t) \nn\\
	&& + \sigma_0 \sum_{k'=1, \text{odd}}^{\min(q,m')}  C_{m'+q-2k'}^{m'-k'} C_{N-(m'+q-2k')}^{k'} p_{m,m'+q-2k'}(t)  \nn\\
	&& + \sigma_1 \sum_{k=0}^{\min(q/2,m)} \sum_{k'=0}^{\min(q/2,m')} \frac{1-(-1)^{k+k'}}2 C_{m+q/2-2k}^{m-k} C_{N-(m+q/2-2k)}^k   \nn\\
	&& \times C_{m'+q/2-2k'}^{m'-k'} C_{N-(m'+q/2-2k')}^{k'}  p_{m+q/2-2k,m'+q/2-2k'}(t) \Big].
\eea

\bibliographystyle{jhep}
\bibliography{references}

\end{document}